\documentclass[12pt]{article}
\usepackage{graphicx,color}
\usepackage{amsmath}
\usepackage{natbib}
\usepackage{geometry}
\usepackage{units}
\usepackage[noline,boxed]{algorithm2e}
\usepackage{caption}

\newcommand{\blind}{0}

\def\C {\,|\:}
\def\x {\bf x}

\begin{document}

\if0\blind
{
	\title{Heteroscedastic BART Via Multiplicative Regression Trees}	
	\author{
	M.~T.~Pratola\thanks{Department of Statistics, The Ohio State University, 1958 Neil Avenue, 404 Cockins Hall, Columbus, OH 43210-1247 (mpratola@stat.osu.edu).},
	H.~A.~Chipman\thanks{Department of Statistics, Acadia University.},
	E.~I.~George\thanks{Department of Statistics, The Wharton School,  University of Pennsylvania.},
	and
	R.~E.~McCulloch\thanks{School of Mathematical and Statistical Sciences, Arizona State University.}
	}
} \fi

\if1\blind
{
  \bigskip
  \bigskip
  \bigskip
	\title{Heteroscedastic BART Via Multiplicative Regression Trees}	
  \medskip
} \fi

\maketitle
\thispagestyle{empty}

\begin{abstract}
{ BART (Bayesian Additive Regression Trees) has become increasingly 
popular as a flexible and scalable nonparametric regression approach for 
modern applied statistics problems.  For the practitioner dealing with large and complex 
nonlinear response surfaces, its advantages include a matrix-free formulation 
and the lack of a requirement to prespecify a confining regression basis.  
Although flexible in fitting the mean, BART has been limited by its reliance on
a constant variance error model.
This homoscedastic assumption is unrealistic in many applications. 
Alleviating this limitation, we propose HBART, a nonparametric heteroscedastic elaboration of BART. 
In BART, the mean function is modeled with a sum of trees,
each of which determines an additive contribution to the mean.
In HBART, the variance function is further modeled with a product  
of trees, each of which determines a multiplicative contribution to the variance.  
Like the mean model, this flexible, multidimensional variance model is entirely nonparametric with no need for the prespecification of a confining basis.  Moreover, with this enhancement, HBART can provide insights into the potential relationships of the predictors with both the mean and  the variance. Practical implementations of HBART with revealing new diagnostic plots are demonstrated with simulated and real data on used car prices, 
fishing catch production and alcohol consumption.}
\\

\end{abstract}

\newpage

\setcounter{page}{1}


\section{Introduction}
\label{section:intro}

Statisticians increasingly face the challenging task of developing flexible regression approaches for the discovery and estimation of important relationships between a high-dimensional vector of potential predictors ${\x}$ and a continuous response $Y$, and for subsequent inferences such as prediction with full uncertainty quantification.  A number of popular methods emerging in this direction include dimension-reduction techniques 
\citep{Allen:etal:2013, Cook:2007},  
statistical model approximations \citep{Sang:Huang:2012}, 
machine learning techniques such as boosting
\citep{Freund:Schapire:1997, Freidman:2001}, 
bagging and random forests 
\citep{Breiman:2001}, 
and Bayesian tree models, 
\citep{chipman:etal:2002, Gramacy:Lee:2008, chipman:etal:2010, Taddy:etal:2011}.  An advantage of the Bayesian approaches is their inherent ability to provide  full posterior inference, in particular their automatic quantification of predictive uncertainty with a predictive distribution. 

With an emphasis on point prediction, many of these developments have been limited to modeling the relationship between $Y$ and $\x$ with a conditional mean function $E[Y \C \x]$ representation plus a constant variance error term.  This is tantamount to assuming that the conditional variance function $Var[Y \C \x] \equiv$ $Var[Y]$ is independent of $\x$ and constant.   Such a homoscedastic assumption is unrealistic in many practical applications, and when violated can lead to miscalibrated predictive uncertainty quantification as well as the overfitting of $E[Y \C \x]$.  Furthermore, when heteroscedasticity is present, the actual variance function $Var[Y \C \x]$ may capture important relationships between $\x$ and the variance of $Y$, which can in themselves be of fundamental interest.

In high-dimensional scenarios, 
the ability to flexibly account for heteroscedasticity and 
explore the relationship between variance and predictors is, 
to the best of our knowledge, very limited.  
Yet there is broad recognition of the importance of this problem.  
As made clear by  \cite{daye:etal:2012}, the problem of 
heteroscedasticity has largely been ignored in the analysis of 
high-dimensional genomic data even though it is known to be a common feature of biological data. 
They address this problem with basis expansions of the mean and log variance as in \cite{carroll:1988}, 
introducing sparsity with penalized likelihood estimation similar to the 
LASSO \citep{tibshirani:1996}.  
However, this already challenging framework for uncertainty quantification relies on the prespecification of a useful regression basis, 
which may not be feasible in high-dimensional 
settings.
\cite{goldberg1998regression} take a log-transformation approach to arrive at a Gaussian Process model with latent variance parameters modeled as a mixture of Gaussians, a method limited to small datasets due to the need for inversion of a dense covariance matrix.  Note also that while simple transformations of $Y$, such as Box-Cox transformations or the more general Yeo-Johnson transformations \citep{Yeo:Johnson:2000}, can sometimes mitigate heteroscedasticity for classical linear model settings, they are generally too limited in scope to extract the potential information in $Var[Y \C \x]$, especially when $\x$ is high dimensional.  


The main thrust of this paper is to propose HBART, a heteroscedastic elaboration of BART (Bayesian Additive Regression Trees) that accounts for potential heteroscedasticity by simultaneously modeling both $E[Y \C \x]$ and $Var[Y \C \x]$ with two distinct nonparametric multidimensional representations.   Just as in BART \citep{chipman:etal:2010}, HBART continues to model the conditional mean function $E[Y \C \x]$ with a ``sum-of-trees'' ensemble, each tree of which determines an additive contribution to the mean.  However, instead of treating the conditional variance of $Y$ as an unknown constant, HBART further models $Var[Y \C \x]$ with a ``product-of-trees'' ensemble, each tree of which determines a multiplicative contribution to the variance.  
This second ensemble takes advantage of the tree-basis to form a flexible variance response surface defined product-wise. As will be seen, this allows for conditional conjugate variance priors which greatly facilitates computation and prior elicitation.  

The tree ensembles in both BART and HBART are comprised of the adaptive Bayesian regression trees that were introduced to provide the basic framework for Bayesian CART  \citep{chipman:etal:1998, denison:etal:1998} and its elaborations to Bayesian treed models \citep{chipman:etal:2002,Gramacy:Lee:2008,Taddy:etal:2011}.  While these Bayesian single tree model formulations can accommodate heteroscedasticity, the extra flexibility of a tree ensemble offers a richer representation for better function approximation, and is dramatically more hospitable for effective MCMC computation.  An alternative, though less flexible, heteroscedastic elaboration of BART was proposed by \cite{Bleich:Kapelner:2014}, who added a linear parametric model for the variance rather than a nonparametric ensemble.  In another interesting direction, \cite{Murray:2017} proposed a log-linear additive tree model admitting some forms of heteroscedasticity with a development focused on modeling categorical and count data.

\begin{figure}[ht!]
\begin{center}
\includegraphics[scale=.5]{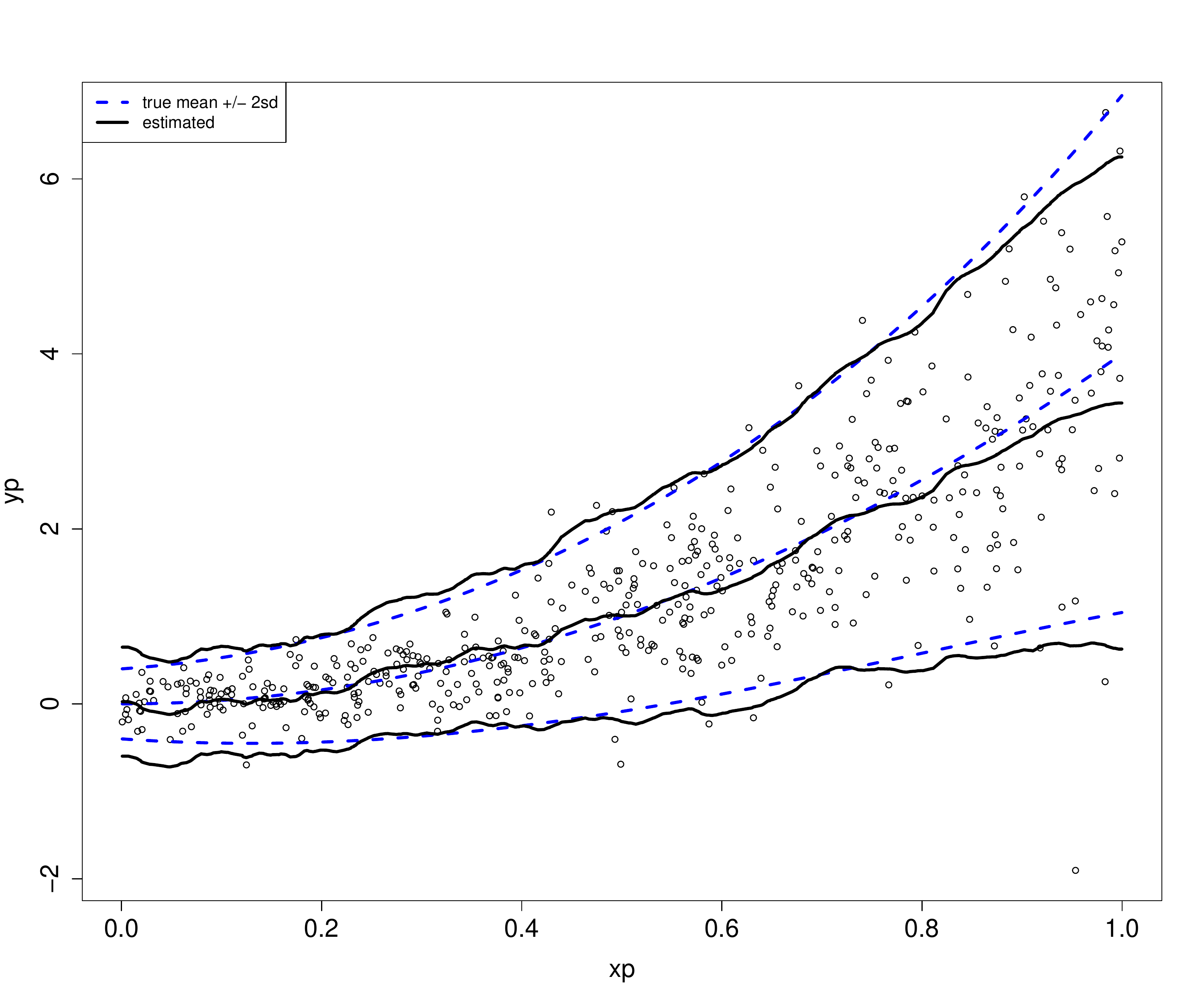}
\end{center}
\caption{
Simulated example.
The underlying true model is depicted by the dashed curves.
The middle dashed curve is $f(x) = E[Y \C \x]$ and the upper and lower
dashed curves are drawn at $f(x) \pm 2 \, s(x)$
where $s(x) = \sqrt{Var[Y \C \x]}$.
The middle solid curve is $\hat{f}(x)$ and the upper and lower
solid curves are drawn at $\hat{f}(x) \pm 2 \, \hat{s}(x)$
where $\hat{f}(x)$ and $\hat{s}(x)$ are HBART estimates. 
The symbols are the simulated $(x_i,y_i)$ pairs.
}\label{fig:oneDsim-ests}
\end{figure}

As a first example to motivate our modeling scenario of interest, and which is further discussed in Section \ref{section:simulated}, consider the simple single predictor simulated dataset  shown in Figure \ref{fig:oneDsim-ests}.  This figure allows us to easily visualize the HBART estimates for this data.  First, the true mean function and plus-or-minus twice the true predictor-dependent error standard deviation are shown using dashed lines.  Second, the HBART estimates of these three curves are displayed with solid lines.  Figure \ref{fig:oneDsim-ests} displays a basic goal of the paper:
we seek to flexibly estimate both the predictor-dependent mean function 
{\it and}  the predictor-dependent variance function. 

In this first example, we deliberately used a one-dimensional predictor space to facilitate  transparent visualization.   Of course, it would not be difficult to come up with standard parametric and nonparametric methods that would perform just as well on this simple example.  However, as the dimension of $\x$ grows larger, extracting information from the regression problem in full generality quickly becomes much more challenging.   To appreciate this, we demonstrate HBART on three more challenging multidimensional real examples in Section \ref{section:examples}.  In Section \ref{section:cars}, we consider a data set of used car prices with 15 potential predictors, and present an extended data analysis illustrating the rich inferential possibilities offered by HBART.  In Section \ref{section:Fish-Alc} we present two additional examples, one on fishing catch production with 25 potential predictors, and the other on alcohol consumption with 35 potential predictors.  Throughout these examples, we also introduce and illustrate new graphical tools for diagnosing the extent of heteroscedasticity and for gauging model fit.

The paper proceeds as follows.  
In Section \ref{section:model} we review the original homoscedastic version of BART. 
In Section \ref{section:hetmodel}, 
we develop the new heteroscedastic HBART model 
which builds on BART's additive representation for the mean 
component with the addition of a multiplicative component for the variance model.  
The potential of HBART along with our new diagnostic tools is illustrated on four examples in Section \ref{section:examples}.  Finally, we conclude with a discussion in  Section \ref{section:conc}.

\section{The Homoscedastic BART Model}\label{section:model}
\label{section:bart}
For high-dimensional regression, most statistical  and machine learning techniques focus on the estimation of $E[Y\C\x]=f(\x)$.
When a model for conditional variance is explicitly considered, it is typically assumed that $Var[Y \C \x]=\sigma^2$ with
the data generated according to 
\begin{align}
\label{eq:homoprocess}
Y({\bf x}) = f({\bf x})+\sigma Z
\end{align}
where $Z\sim\text{N}(0,1)$ and ${\bf x}=(x_1,\ldots,x_d)$ is a $d$-dimensional vector of predictor variables.  Such a process is known as a homoscedastic process.  

\begin{figure}[ht!]
\begin{center}
\includegraphics[scale=.5]{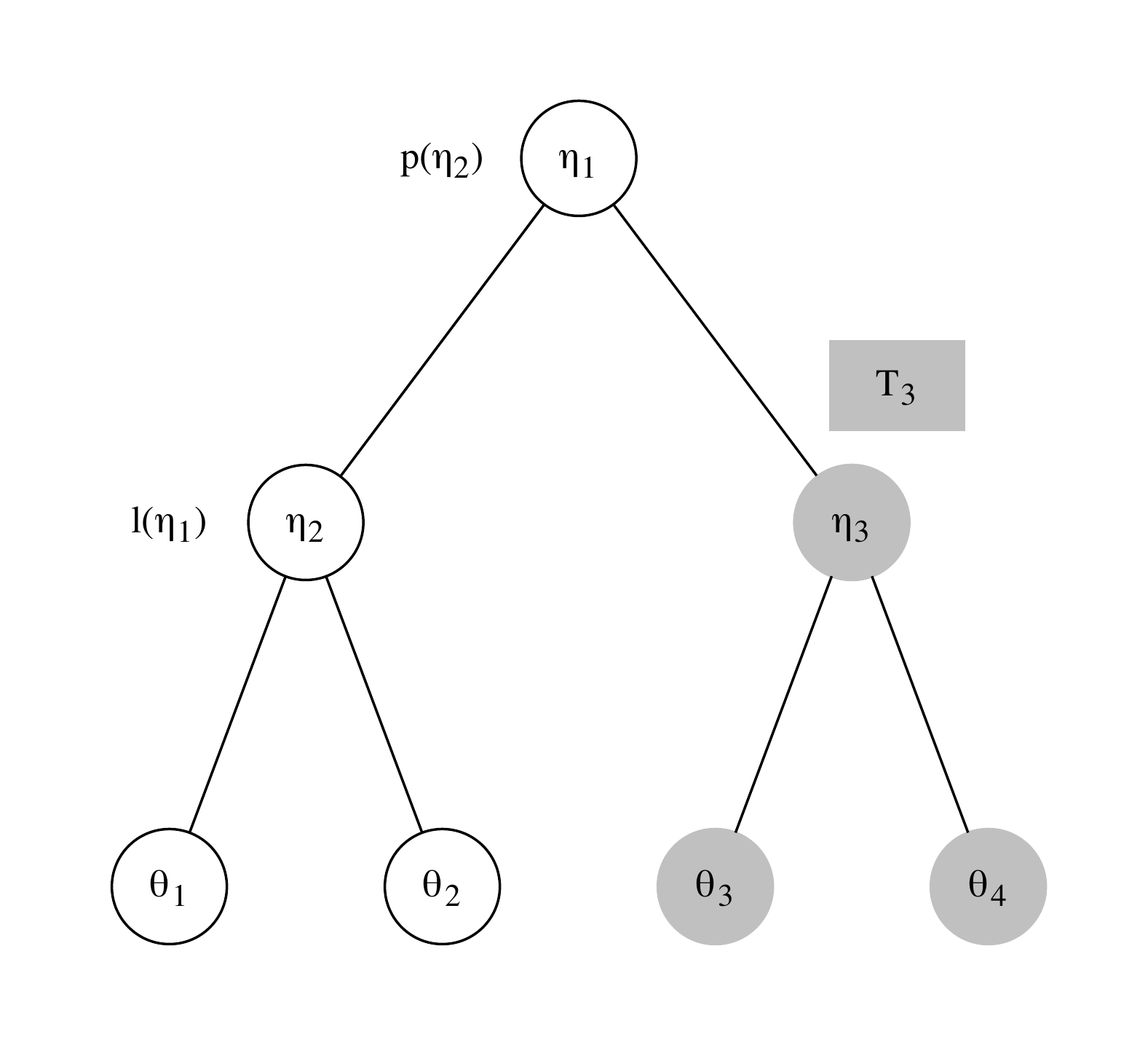}
\end{center}
\caption{Labeling for a single regression tree ${\bf T}$.  Nodes are denoted by circles and labeled using the symbol $\eta$.  Lines denote branches connecting the nodes.  Nodes can also be identified as left and right children (e.g. $\eta_2=l(\eta_1)$) or as parent (e.g. $\eta_1=p(\eta_2)$).  Terminal nodes have no branches below them and contain associated parameter values $\theta$.  A sub-tree uses the $T$ symbol with subscript given by the node index (e.g. the subtree including $\eta_3$ and all its children is $T_3$).  Note that later in the paper ${\bf T}_j$ will also index one member of an ensemble of trees, its use will be clear from context.}
\label{fig:treelabels}
\end{figure}

 
As previously described, BART models the unknown mean function $f(\x)$ with an ensemble of Bayesian regression trees.  Such regression trees provide a simple yet powerful non-parametric specification of a multidimensional regression bases, where the form of the basis elements are themselves learned from the observed data.
Each Bayesian regression tree is a recursive binary tree partition that is made up of interior nodes, ${\bf T}$, and a set of parameter values, ${\bf M}$, associated with the terminal nodes.  Each interior tree node, $\eta_i$, has a left and right child, denoted $l(\eta_i)$ and $r(\eta_i)$.  In addition, all nodes also have one parent node, $p(\eta_i),$ except for the tree root.  One may also refer to a node by a unique integer identifier $i$, counting from the root where the left child node is $\eta_{2i}$ and the right child node is $\eta_{2i+1}$.  For example, the root node $\eta_1$ is node $1$.  One can also label a subtree starting at node $\eta_i$ simply as $T_i$.  Figure \ref{fig:treelabels} summarizes our notation.

Internal nodes of regression trees have split rules depending on the predictors and ``cutpoints'' that are the particular predictor values at which the internal nodes split.  This modeling structure is encoded in ${\bf T}$, which accounts for the split rules at each internal node of a tree and the topological arrangement of nodes and edges forming the tree.  Given the design matrix ${\bf X}$ of predictors having dimension $n\times d$, each column represents a predictor variable $v, v=1,\ldots,d$ and each row ${\bf x}$ corresponds to the observed settings of these predictors. At a given internal node, the split rule is then of the form $x_v<c$ where $x_v$ is the chosen split variable and $c$ is the chosen cutpoint $c$ for split variable $x_v.$

The Bayesian formulation proceeds by specifying discrete probability distributions 
on the split variables $v$ taking on a value in $\{1,\ldots,d\}$ and specifying 
discrete probability distributions on the set of distinct possible cutpoint values, 
where $n_v$  is the total number of discrete cutpoints available for variable $v$.  
For a continuous predictor, a choice of $n_v=100$ is common \citep{chipman:etal:2010}.  
Categorical predictors are handled by creating binary dummy indicator variables for each category
(not all but one as in the linear model case)
and then using a single cutpoint.  
An interesting third type of variable is the ordinal categorical variable.
If we label $C$ ordinal levels with $1,2,\ldots,C$ then it might be reasonable to use $nv=C-1$ cutpoints
at $i+.5, \; i=1,2,\ldots (C-1)$. 
The internal modeling structure of a tree, ${\bf T},$ can then be expressed as ${\bf T}=\{(v_1,c_1),(v_2,c_2),\ldots\}$.

The Bayesian formulation is completed by specifying prior distributions on the parameters at the terminal nodes. For $n^g=|{\bf M}|$ terminal nodes in a given tree, the corresponding parameters are ${\bf M}=\{\theta_1,\ldots,\theta_{n^g}\}$. Taken all together, the Bayesian regression tree defines a function $g({\bf x};{\bf T},{\bf M})$ which maps input ${\bf x}$ to a particular $\theta_j,$ $j\in 1\ldots n^g.$


The original BART model is then obtained as the ensemble sum of $m$ such Bayesian regression trees plus a constant variance error component, 
\[
y({\x}_i) = \sum_{j=1}^m g({\x}_i;{\bf T}_j,{\bf M}_j) + \sigma Z_i, \ \ Z_i \sim N(0,1)
\]
where $y({\x}_i)$ is the observation collected at predictor setting $\x_i$, and $\sigma^2$ is the variance of the homoscedatic process.  Since the trees are modeling the components of the mean response in BART, one usually labels the bottom node parameters as $\theta\equiv \mu$.  Then, normal priors are specified for the bottom node parameters
\[
\pi(\mu_{jk}) \sim N(0,\tau^2),
\]
where $\mu_{jk}$ is the $k$th bottom node component for tree $j$, along with an inverse chi-squared prior for the variance,
\[
\sigma^2 \sim \chi^{-2}(\nu,\lambda).
\]
where 
$\chi^{-2}(\nu,\lambda)$ denotes the distribution
$(\nu \lambda)/\chi^2_\nu$.

For a prior on the tree structure, we specify a stochastic process that describes how a tree is drawn.
A node at depth $d$ spawns children with probability
\[
\alpha(1+d)^{-\beta},
\]
for $\alpha\in (0,1)$ and $\beta\geq 1.$
As the tree grows, $d$ gets bigger so that a node is less likely to spawn children and
more likely to be remain a bottom node, thus penalizing tree complexity.
Details on specifying the parameters of the prior distributions are discussed 
in detail in \cite{chipman:etal:2010}, while typically the choice $m=200$ 
trees appears to be reasonable in many situations.

The use of normal priors on the terminal node $\mu$'s and an inverse chi-square prior on the constant variance, greatly facilitates the posterior simulation via our MCMC algorithm as they are conditionally conjugate.
  Selecting the split variables and cutpoints of internal tree nodes is performed using a Metropolis-Hastings algorithm by growing and pruning each regression tree within a sequential Bayesian backfitting algorithm. The growing/pruning are performed by birth and death proposals which either split a current terminal node in ${\bf M}$ on some variable $v$ at some cutpoint $c$, or collapse two terminal nodes in ${\bf M}$ to remove a split.  For complete details of the MCMC algorithm, the reader is referred to \cite{chipman:etal:1998, denison:etal:1998,chipman:etal:2010,Pratola:2016}.

\section{The Heteroscedastic HBART Model}\label{section:hetmodel}

As suggested earlier, real world data does not always follow the simple constant-variance process modeled by BART.   Instead, it may be of more interest to model a process $Y({\bf x})$ that is formed by an unknown mean function, $E[Y\vert {\bf x}]=f({\bf x}),$ {\em and} an unknown variance function, $Var[Y\vert {\bf x}]=s^2({\bf x}),$ along with a stochastic component arising from independent random perturbations $Z.$  This process is assumed to be generated according to the relation
\begin{align}
\label{eq:heteroprocess}
Y({\bf x}) = f({\bf x})+s({\bf x})Z
\end{align}
where again $Z\sim\text{N}(0,1)$ and ${\bf x}=(x_1,\ldots,x_d)$ is a $d$-dimensional vector of predictor variables.  For simplicity of exposition, these predictor variables are assumed to be common to both $f(\x)$ and $s(\x),$ although this is not strictly necessary in the proposed approach.  This heteroscedastic process is the natural generalization of the homoscedastic ``Normal-errors with constant variance'' assumption of (\ref{eq:homoprocess}) that is pervasive in the statistics and machine learning literatures.  The types of inference in this setting are broader in scope than the usual case.  That is, in addition to predicting the mean behavior of the process by estimating $f({\bf x})$ and investigating the importance of predictor variables on the mean response, one is also interested in inferring the variability of the process by estimating $s({\bf x})$ and investigating which predictors are related to the process variability.  

Under data for which \eqref{eq:heteroprocess} is appropriate, our proposed HBART methodology models the unknown mean function, $f({\bf x}),$ and the unknown variance function, $s^2({\bf x}),$ using two distinct ensembles of Bayesian regression trees: an additive regression tree model for the mean as in BART, $$f({\bf x}) = \sum_{j=1}^mg({\bf x};{\bf T}_j,{\bf M}_j),$$ and a multiplicative regression tree model for the variance component, 
\begin{align}
s^2({\bf x}) &= \prod_{l=1}^{m^\prime}h({\bf x};{\bf T}^\prime_l,{\bf M}^\prime_l).
\label{eq:vardecomp}
\end{align} 
In this model, ${\bf T}_j$ encodes the structure of the $j^{th}$ tree for the {\em mean} and ${\bf M}_j=\lbrace\mu_{j,1},\ldots,\mu_{j,n_j^g}\rbrace$ are the $n_j^g=|{\bf M}_j|$ scalar terminal-node parameters for the mean in each tree.  Similarly, ${\bf T}^{\prime}_l$ encodes the structure of the $l^{th}$ tree for the {\em variance} and in this case we represent the bottom node maps as $\theta\equiv s^2$ so that ${\bf M}^{\prime}_l=\lbrace s^2_{l,1},\ldots,s^2_{l,n_l^h}\rbrace$ are the $n_l^h=|{\bf M}^{\prime}_l|$ scalar  terminal-node parameters for the variance in each tree.
In other words, $s^2({\bf x}_i)$ is modeled as a product of Bayesian regression trees.  

The posterior of the HBART model can be factored as
\[
\pi({\bf T},{\bf M},{\bf T}^\prime,{\bf M}^\prime\vert {\bf y},{\bf X}) \propto L({\bf y}|{\bf T},{\bf M},{\bf T}^\prime,{\bf M}^\prime,{\bf X})\prod_{j=1}^m\pi({\bf T}_j)\pi({\bf M}_j|{\bf T}_j)\prod_{l=1}^{m^\prime}\pi({\bf T}_l^\prime)\pi({\bf M}_l^\prime|{\bf T}_l^\prime)
\]
where
\[\pi({\bf M}_j|{\bf T}_j)=\prod_{k=1}^{n^g}\pi(\mu_{jk})\]
and 
\[\pi({\bf M}^\prime_l|{\bf T}^\prime_l)=\prod_{k=1}^{n^h}\pi(s^2_{lk}).\]
This specification assumes a priori independence of terminal node parameters for both the mean and variance, as well as independence of the mean trees and  variance trees.  This approach allows for easy specification of priors for the mean and variance model components and straightforward use of conditional conjugacy in implementing the MCMC sampler, which eases computations.

The proposed heteroscedastic regression tree model outlined above is fitted using a Markov Chain Monte Carlo (MCMC) algorithm.  
The basic steps of the algorithm are given in Algorithm \ref{alg:mcmc}.
The algorithm is a Gibbs sampler in which we draw each 
$({\bf T}_j,{\bf M}_j)$  and 
$({\bf T}^\prime_j,{\bf M}^\prime_j)$  
conditional on all other parameters and the data.
To draw 
$({\bf T}_j,{\bf M}_j)$  we integrate out ${\bf M}_j$
and draw ${\bf T}_j$ and then ${\bf M}_j \vert {\bf T}_j$.
The conditionally conjugate prior specifications
in Section \ref{subsec:mean-model} allow us to do the
integral and draw easily.
Our product of trees model and prior specifications
(Section \ref{subsection:variance-model})
allow us to use the same strategy to draw 
$({\bf T}^\prime_j,{\bf M}^\prime_j)$.
The draws of ${\bf T}_j \vert \cdot$ and ${\bf T}_j^\prime \vert \cdot$
are done using Metropolis-Hastings steps as in
\cite{chipman:etal:2010}
and
\cite{Pratola:2016}.

\vspace{.2in}

\begin{algorithm}[H]
\label{alg:mcmc}
 \caption{MCMC steps for the proposed HBART model.}
     \SetAlgoLined
     \KwData{$y_1,\ldots,y_N;\ \ {\bf x}_1,\ldots,{\bf x}_N$}
     \KwResult{Approximate posterior samples drawn from $\pi({\bf T},{\bf M},{\bf T}^\prime,{\bf M}^\prime\vert y_1,\ldots,y_N; {\bf x}_1,\ldots,{\bf x}_N)$ }
     \For{ $N_{mcmc}$ iterations}{
     \For{$j=1,\ldots,m$}{
      i. Draw ${\bf T}_j|\cdot$ \\ 
      ii. Draw ${\bf M}_j|{\bf T}_j,\cdot$ 
     }
     \For{$j=1,\ldots,m^\prime$}{
      iii. Draw ${\bf T}^\prime_j|\cdot$ \\ 
      iv. Draw ${\bf M}^\prime_j|{\bf T}^\prime_j,\cdot$ 
     }
     }
\end{algorithm}

Next we outline the mean and variance models in detail as well as the full conditionals required for implementing Algorithm \ref{alg:mcmc}.

\subsection{Mean Model}\label{subsec:mean-model}
Viewed as a function of $\mu_{jk}$, the HBART likelihood in the $k^{th}$ terminal node of the $j^{th}$ mean tree is

\[
L(\mu_{jk}|\cdot)=\prod_{i=1}^n\frac{1}{\sqrt{2\pi}s({\bf x}_i)}exp\left(\frac{(r_i-\mu_{jk})^2}{s^2({\bf x}_i)}\right)
\]

where $n$ is the number of observations mapping to the particular terminal node and 

\[
r_i=y_i-\sum_{q\neq j}g({\bf x}_i;{\bf T}_q,{\bf M}_q).
\]

As in BART, the conjugate prior distribution for the mean component is

\[
\pi(\mu_{jk})\sim N(0,\tau^2), \ \forall j,k.
\]

Then the full conditional for the mean component is

\begin{align}
\pi(\mu_{jk}|\cdot) \sim N\left(\frac{\sum_{i=1}^n\frac{r_i}{s^2({\bf x}_i)}}{\frac{1}{\tau^2}+\sum_{i=1}^n\frac{1}{s^2({\bf x}_i)}},\frac{1}{\frac{1}{\tau^2}+\sum_{i=1}^n\frac{1}{s^2({\bf x}_i)}}\right)
\label{eq:meanupdate}
\end{align}

and (ignoring terms which cancel) the integrated likelihood is

\begin{align}
\int L(\mu_{jk}|\cdot)\pi(\mu_{jk})d\mu_{jk}\propto\left(\tau^2\sum_{i=1}^n\frac{1}{s^2({\bf x}_i)}+1\right)^{-1/2} \, exp\left(\frac{\frac{\tau^2}{2}\left(\sum_{i=1}^n\frac{r_i}{s^2({\bf x}_i)}\right)^2}{\tau^2\sum_{i=1}^n\frac{1}{s^2({\bf x}_i)}+1}\right)
\label{eq:mujmarglik}
\end{align}

which depend on the data only via the sufficient statistic

\[
\sum_{i=1}^n\frac{r_i}{s^2({\bf x}_i)}.
\]

These forms are nearly identical to those of the BART model, with the only change arising from replacing a scalar variance $s^2$ with a vector variance $s^2({\bf x}_i), i=1,\ldots,n.$  Conditional on a drawn realization of the variance component at the $n$ observation sites, steps (i) and (ii) of Algorithm \ref{alg:mcmc} are analogous to the BART sampler, where now instead of needing to calculate the sample means of data mapping to each tree's bottom node, one needs to calculate the variance-normalized means of the data to perform the Gibbs update of equation (\ref{eq:meanupdate}) and for updating the mean tree structures using the marginal likelihood of equation (\ref{eq:mujmarglik}).

\subsection{Calibrating the  Mean Prior}\label{subsec:calibrate-mean-prior}

As in BART, the specified prior for $\mu_{jk}$ implies a prior on the mean function, $f({\bf x}) \sim \text{N}(0,m\tau^2).$  This prior assumes mean-centered observations so that the mean of the prior on $\mu_{jk}$ is simply $0$ as shown above.  Calibrating the variance of the prior proceeds using a weakly data-informed approach by taking the minimum and maximum response values from the observed data, $y_{min}, y_{max}$ and assigning a high probability to this interval, i.e. setting
\[\tau=\frac{y_{max}-y_{min}}{2\sqrt{m}\kappa}\]
where, for instance, $\kappa=2$ species a 95\% prior probability that $f({\bf x})$ lies in the interval $(y_{min},y_{max})$ \citep{chipman:etal:2010}.  

Essentially, the hyperparameter $\kappa$ controls the bias-variance tradeoff: the higher is $\kappa$, the greater the probability that the mean function accounts for the range of observed data implying a smaller variance, while the smaller is $\kappa$, the less probability that the mean function accounts for the range of observed data implying a larger variance.  In BART, the recommended default setting is $\kappa=2$, but for HBART we have found that a higher default is preferable, such as $\kappa=5$ or $\kappa=10.$  Ideally, we recommend carefully selecting $\kappa$ in a prescribed manner in practice.  In the examples of Section \ref{section:examples}, a simple graphical approach and a cross-validation approach to selecting $\kappa$ will be explored.

\subsection{Variance Model}\label{subsection:variance-model}
Viewed as a function of $s^2_{lk}$, the HBART likelihood in the $k^{th}$ terminal node of the $l^{th}$ variance tree is

\[
L(s^2_{lk}|\cdot) = \prod_{i=1}^n\frac{1}{\sqrt{2\pi}s_{lk}}exp\left(-\frac{e_i^2}{2s^2_{lk}}\right),
\]

where $n$ is the number of observations mapping to the particular terminal node and  \[e_i^2=\frac{\left(y({\bf x}_i)-\sum_{j=1}^m g({\bf x}_i;{\bf T}_j,{\bf  M}_j)\right)^2}{s^2_{-l}({\bf x}_i)},\]
where
\[s^2_{-l}({\bf x}_i)=\prod_{q\neq l}h({\bf x}_i;{\bf T}^{\prime}_q,{\bf M}^{\prime}_q).\]


We specify a conjugate prior distribution for the variance component as

\[
s_{lk}^2 \sim \chi^{-2}\left(\nu^{\prime},\lambda^{\prime}\right),\ \ \forall l,k. 
\]


Then, it is readily shown that the full conditional for the variance component is
\begin{align}
s_{lk}^2|\cdot &\sim \chi^{-2}\left(\nu^{\prime}+n,\frac{\nu^{\prime}{\lambda^{\prime}}^2+\sum_{i=1}^n\frac{e_i^2}{s^2_{-l}({\bf x}_i)}}{\nu^{\prime}+n}\right)
\label{eq:sigmajupdate}
\end{align}
so that the terminal nodes of the $m^{\prime}$ variance component trees in the product decomposition (\ref{eq:vardecomp}) are easily updated using Gibbs sampler steps.  The integrated likelihood is also available in closed form,

\begin{align}
\int L(s_{lk}^2|\cdot)\pi(s_{lk}^2)ds_{lk}^2 &= \frac{\Gamma(\frac{\nu^{\prime}+n}{2})\left(\frac{\nu^{\prime}{\lambda^{\prime}}^2}{2}\right)^{\nu^{\prime}/2}}{\left(2\pi\right)^{n/2}\prod_{i=1}^ns_{-l}({\bf x}_i)\Gamma(\nu^{\prime}/2)\left(\nu^{\prime}{\lambda^{\prime}}^2+\sum_{i=1}^ne_i^2\right)^{\frac{\nu^{\prime}+n}{2}}}
\label{eq:sigmajmarglik}
\end{align}
which depends on the data only via the sufficient statistic,
\[\sum_{i=1}^ne_i^2.\]
This closed-form solution allows for easily exploring the structure of variance trees ${\bf T}^{\prime}_1,\ldots,{\bf T}^{\prime}_{m^{\prime}}$ using Metropolis-Hastings steps in the same way that the mean model trees are explored.
That is, the conjugate inverse chi-squared prior leads to a Gibbs step drawing from an inverse chi-squared full conditional when updating the components of ${\bf M}^\prime$ using equation (\ref{eq:sigmajupdate}).  Sampling the tree structure ${\bf T}_j^\prime$ is performed via Metropolis-Hastings steps via the marginal likelihood (\ref{eq:sigmajmarglik}). 
This is made possible as the integrated likelihood is analytically tractable with the heteroscedastic model specified.

The interpretation of this model form follows as the mean being factored into a sum of weakly informative components (as usual in BART) while the variance is factored into the product of weakly informative components.  This latter factoring is indexed by the predictor ${\bf x}_i$ where each tree $h({\bf x}_i;{\bf T}^\prime_l,{\bf M}^\prime_l)$ contributes a small component of the variance at ${\bf x}_i$ with the product of the $l=1,\ldots,m^\prime$ trees modeling the overall variance.

For the variance model we again specify discrete uniform priors on the split variables and cutpoints. Note that the number of variance component trees, $m^{\prime},$ need not equal the number of mean component trees, $m.$  A default value that has worked well in the examples explored is $m^\prime=40.$  Since the trees making up the mean model are different from the trees making up the variance model, the number of bottom nodes in the $l$th variance component tree, $n^h_l,$ is unrelated to the number of bottom nodes in the $j$th mean component tree, $n^g_j.$  This means, for instance, that the complexity of the variance function may be different from that of the mean function, and the predictors that are important for the variance function may also differ from those that are important for the mean.

Similar to the mean model component, a penalizing prior is placed on the depth of variance component trees,
with the probability of a node spawning children equal to $\alpha^\prime(1+d)^{-\beta^\prime}$
where $d$ is the depth of the node.  
Typically, the specification of this prior is chosen similar to that used in the mean model components, 
i.e. $\alpha^\prime=0.95, \beta^\prime=2$ specifies a prior preference for shallow trees having a depth of 2 or 3.

\subsection{Calibrating the Variance Prior}\label{subsection:calibrate-variance-prior}

The simplicity of the prior for $f({\bf x})$ (Section \ref{subsec:calibrate-mean-prior})
is a major strength of BART.
Our prior for $s({\bf x})$ is more complicated.
We show in this section that a simple strategy for assessing the prior gives
very reasonable results.

From Section \ref{subsection:variance-model} our prior is
\[
s({\bf x})^2 \sim \prod_{l=1}^{m^\prime}s^2_l, \;\;\; \mbox{  with  } s^2_l \sim \chi^{-2}(\nu^\prime,\lambda^\prime), \, \mbox{ i.i.d.}
\]
As in the case of 
$f({\bf x})$,
the prior for
$s({\bf x})$,
does not depend on ${\bf x}$.

Selecting the prior parameters $\nu^\prime,\lambda^\prime$ for the variance components may be done in the following way.  
We suppose we start with a prior for a single variance in the context of the homoscedastic BART model 
$Y=f({\bf x}) + \sigma Z$
with \[\sigma^2 \sim \chi^{-2}(\nu,\lambda).\]
\cite{chipman:etal:2010} discuss strategies for choosing the parameters $(\nu,\lambda)$ in this case.
We can choose a prior in the heteroscedastic model to match the prior in the homoscedastic case by matching
the prior means.
We have
\[E[\sigma^2]=\frac{\nu\lambda}{\nu-2},\]
and
\[
E[s({\bf x})^2]=\prod_{l=1}^{m^\prime}E[s_l^2] = \lambda^{m^\prime} \; \left( \frac{\nu^\prime}{\nu^\prime -2} \right)^{m^\prime}. 
\] 
We then match the means by separately matching the ``$\lambda$ piece'' and the ``$\nu$ piece'' giving
\[
\lambda^\prime=\lambda^\frac{1}{m^\prime}, \;\;\;\;\;\;
\nu^\prime=\frac{2}{1-\left(1-\frac{2}{\nu}\right)^{1/m^\prime}}.
\]

\begin{figure}[ht!]
\begin{center}
\includegraphics[scale=.6]{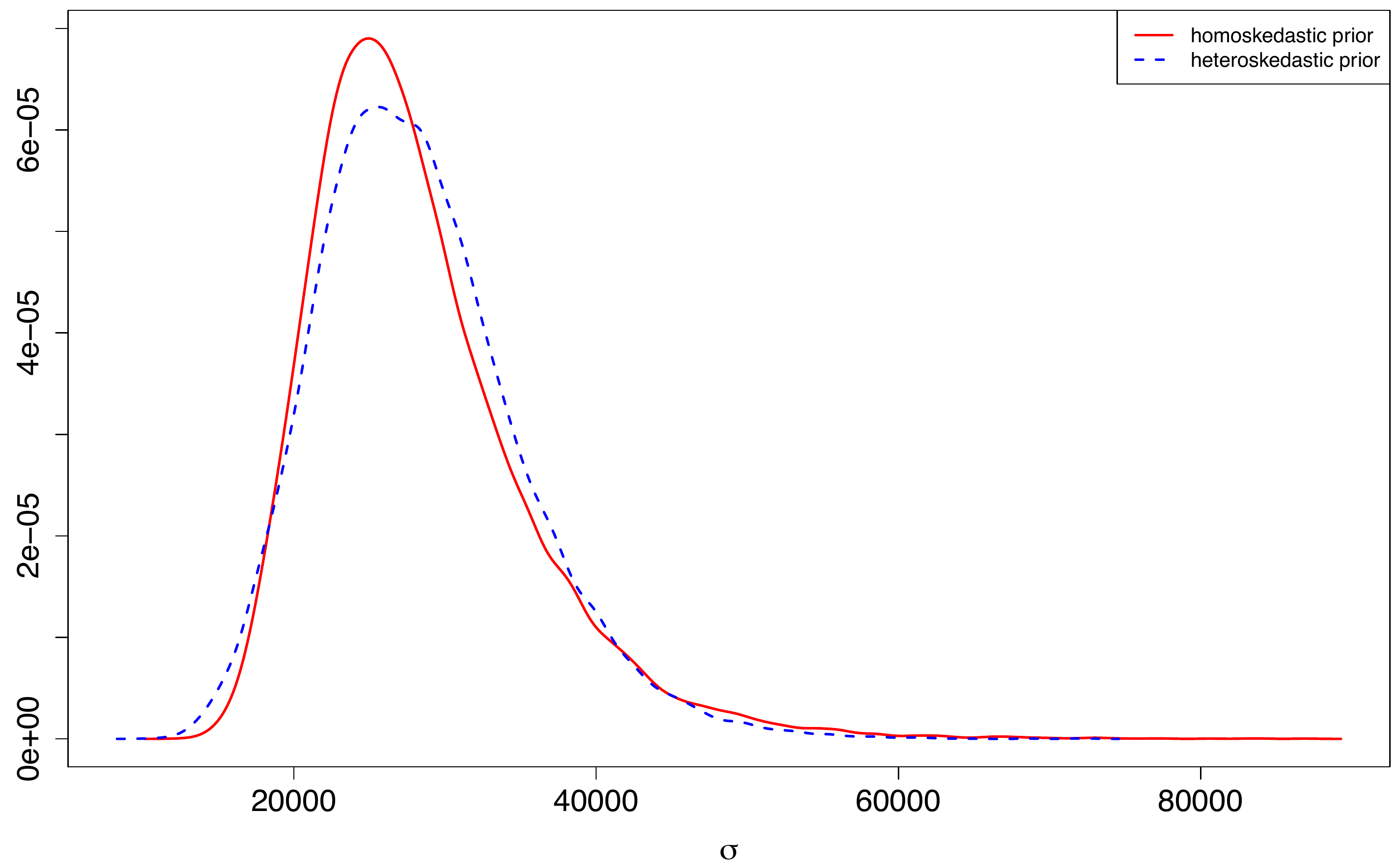}
\end{center}
\caption{
Choosing the prior for the error variance.
The plot is on the standard deviation scale.
The solid line is a prior choice for $\sigma$.
The dashed line is the (approximate) density curve for a matching prior for $s({\bf x})$ calculated via simulation.
}\label{fig:checkprior}
\end{figure}

Figure \ref{fig:checkprior} illustrates this procedure.
For a problem in which the response is the price of (very expensive)
used cars we elicited a prior on $\sigma$ with $\nu=10$ and $\lambda = 26000^2$.
The resulting prior density for $\sigma$ is plotted with a solid line.  Using the formulas above with $m^\prime = 40$ we get $\nu^\prime = 360$ and $\lambda^\prime =  1.66$.
The resulting prior density for $s({\bf x})$ is plotted with a dashed line.  In both cases, we simply took a large number of draws from the prior and then reported the density
smooth of the draws.
These priors are remarkably similar both in their location {\it and} in their general shape.
Using this approach, prior specification is no more complicated that in the homoscedastic case.

%
%

\section{HBART in Action}\label{section:examples}
We now demonstrate the proposed HBART methodology on several examples. 
Our first example 
(Section \ref{section:simulated}) 
is the simulated example with a  one dimensional $\x$, which we introduced in Section \ref{section:intro}.
As we saw, simple graphics can be used to transparently display the results with just a single predictor.
The second example
(Section \ref{section:cars})
is an extended real example in which we develop a  model to predict the price
of a used car using car characteristics.
Graphical examination of the data shows that both nonlinearity and heteroscedasticity are present.
Our model successfully captures both of these. 
We then briefly present two real examples
(Section \ref{section:Fish-Alc})
with {\bf x} having dimension 25 and 35.
In these examples, the basic $Y=f({\bf x}) + s({\bf x}) Z$ structure conflicts a bit with the nature of $Y$, but we show that
our model nevertheless finds a reasonable approximation that is relatively interpretable.

In all our examples we introduce two new graphical diagnostic plots to visualize the effectiveness of HBART
and compare it to the performance of BART.
In the {\it H-evidence} plot we display posterior intervals for $s({\bf x}_i)$ sorted by the values
of $\hat{s}({\bf x}_i)$.  As such, the H-evidence plot is a simple graphic to detect evidence in favor of heteroscedasticity.
Using this plot we can quickly see if the conditional variance is predictor dependent.
In the {\it predictive qq-plot} we start with a sample of observations $({\bf x}_i,y_i)$ (in or out of sample).
For each ${\bf x}_i$ we compute quantiles of $y_i$ based on the predictive distribution $Y \vert {\bf x}_i$ obtained
from our model.  If the model  is correct, the quantiles should look like draws from the uniform distribution.
We use qq-plots to compare our quantiles to uniform draws.

In the cars example we demonstrate the use of cross-validation to choose prior settings.
Typically, cross-validation involves the choice of a simple measure of predictive performance such
as RMSE (root mean square error).
However, the goal of HBART is to get a better feeling for the conditional uncertainty
than one typically obtains from statistical/machine-learning methodologies.
For this purpose, we use the {\it $e$-statistic} \citep{Szekely:Rizzo:2004} as a measure of quality of the predictive qq-plot for our target in
assessing out-of-sample performance.

We use the following model and prior specifications unless otherwise stated.
For the number of trees in our ensembles, $m=200$ for the mean model and $m^\prime = 40$ for the variance model.
The tree prior uses $\alpha = 2$ and $\beta = .95$ for both $\lbrace{\bf T}_j\rbrace,$ the mean trees, and $\lbrace{\bf T}^\prime_j\rbrace,$ the variance trees.
The mean prior parameter is $\kappa=2$ with the consequent choice for $\tau$ discussed in Section \ref{subsec:calibrate-mean-prior}.
We use $\nu=10$ and $\lambda$ equal to the sample variance of $y$ to specify a $\sigma$ prior and then use the
approach of Section \ref{subsection:calibrate-variance-prior} to specify the $\nu^\prime$ and $\lambda^\prime$ for the
heteroscedastic $s$ prior.


\subsection{A Simple Simulated Example}\label{section:simulated}

%

We simulated 500 observations from the model
$$
Y_i = 4  x^2 + .2 \, e^{2x} \, Z_i.
$$
so that $Y_i = f(x_i) + s(x_i) \, Z_i$ with $f(x) = 4  x^2$ and $s(x) = .2 \, e^{2x}$.
Each $x$ is drawn independently from the uniform distribution on (0,1)
and each $Z_i$ is drawn independently from the standard normal distribution.
We then simulated an independent data set in the exact same to serve as out-of-sample data.

As we saw in Section \ref{section:intro}, the out-of-sample simulated data, 
simulation model, 
and point estimates are depicted in 
Figure \ref{fig:oneDsim-ests}.
The dashed curves represent the model with the center line being $x$ vs. $f(x)$
and the two outer lines are drawn at $f(x) \; \pm \; 2 \, s(x)$.
The points are plotted at the simulated $(x_i,y_i)$ pairs.

We ran the MCMC for 1,000 burn-in draws and kept 2,000 subsequent draws to represent
the posterior. 
The solid lines in 
Figure \ref{fig:oneDsim-ests}
represent estimates of $\hat{f}(x)$ and $\hat{f}(x) \pm 2 \hat{s}(x)$ 
obtained by averaging 
$\{f_j(x)\}$ 
and 
$\{s_j(x)\}$ 
where $j$ indexes post burn-in MCMC draws.
Although very similar results are obtained using the default $\kappa=2$, we used
$\kappa=5$.  This is appropriate given the very smooth nature of the true $f$.

\begin{figure}[ht!]
\begin{center}
\includegraphics[scale=.6]{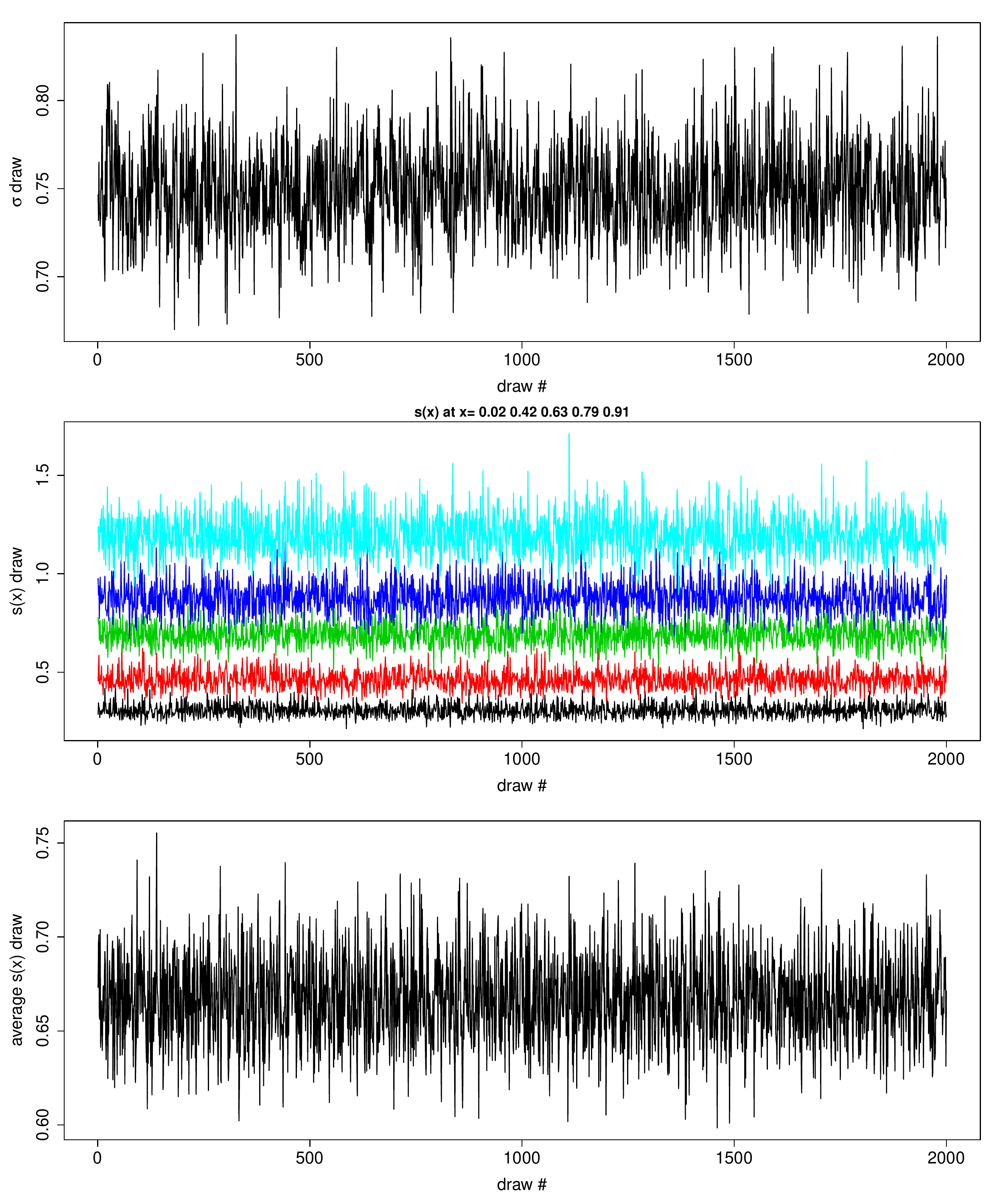}
\end{center}
\caption{
Simulated example.
Top panel: MCMC draws of $\sigma$ in homoscedastic BART.
Middle panel: MCMC draws of $s(x)$ for five different $x$ in HBART.
Bottom panel: MCMC draws of $\bar{s}$ the average of $s(x_i)$ for each MCMC draw in HBART.
}\label{fig:assess-burn-in}
\end{figure}

Figure \ref{fig:assess-burn-in}
informally examines the performance of the MCMC chain by displaying  sequences
of post burn-in draws for certain marginals.
The top panel displays the draws of $\sigma$ from the homoscedastic BART
model $Y = f(x) + \sigma Z$.  For BART, this plot is a simple way
to get a feeling for the performance of the MCMC.  We can see that the draws vary
about a fixed level with an appreciable but moderate level of autocorrelation.
For HBART, there is no simple summary of the overall error level
comparable to the homoscedastic $\sigma$.
The middle panel plots draws of $s(x)$ for 
$x= .12, .42, .63, .79, .91$.
The bottom panel plots the  draws of
$\bar{s} = \frac{1}{n} \sum s(x_i)$,
the average $s$ value for each MCMC draw of $s$.
Here the $x_i$ are from the test data.
In all plots, the MCMC appears to be reasonably well behaved.

\begin{figure}[ht!]
\hspace*{-.2in}\includegraphics[scale=.65]{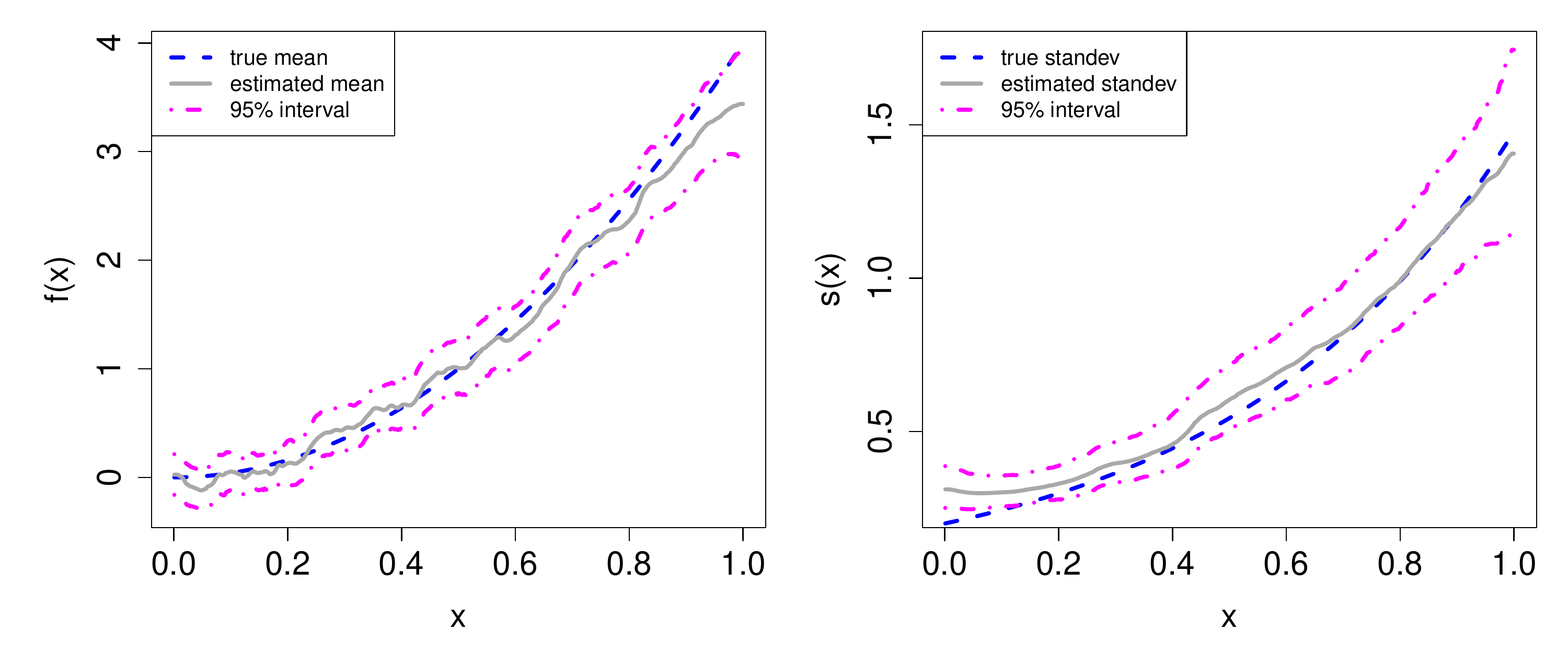}
\caption{
Simulated example.
The left panel displays inference for $f$ and the right panel displays inference for $s$.
In each panel the dashed line is the true function  and the solid line is the estimated function.
The dot-dash lines give point-wise 95\% posterior intervals for $f$ and $s$. 
}\label{fig:oneDsim-ints}
\end{figure}

Figure \ref{fig:oneDsim-ints} displays the uncertainty associated with
our inference for $f$ and $s$.
The left panel displays inference for $f$ and the right panel for $s$.
In each panel, the dashed line is the true function,
the solid line is the estimated function
(the posterior mean estimated by the average of MCMC draws)
and the dot-dash line represents point-wise 95\% posterior intervals
for each $f(x_i)$ (left panel) and $s(x_i)$ (right panel).
The intervals are estimated by the quantiles of the MCMC
draws of $\{f(x_i)\}$ and $\{s(x_i)\}$
and the $x_i$ are from the test data.

\begin{figure}[ht!]
\begin{center}
\includegraphics[scale=.425]{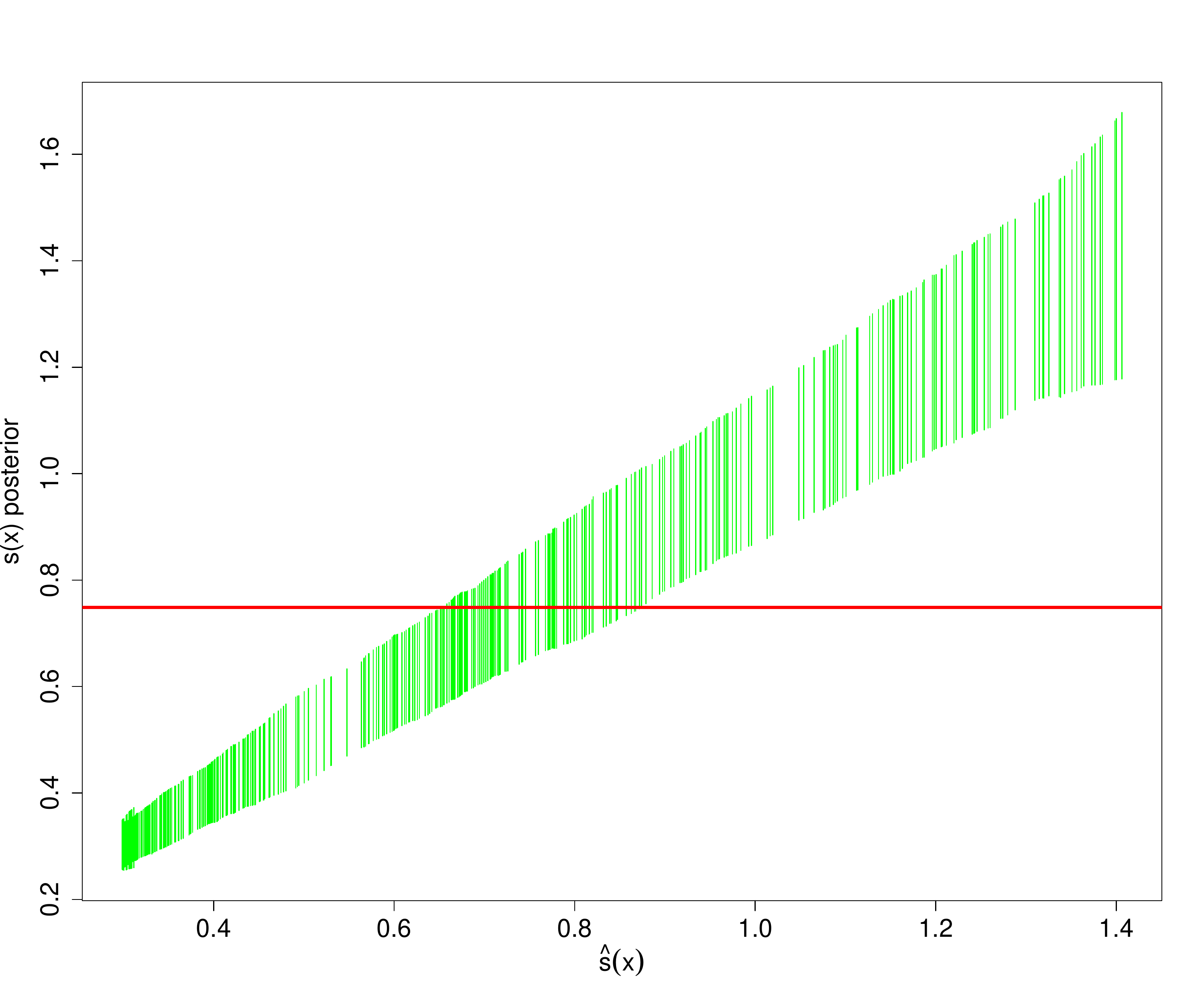}
\end{center}
\caption{
Simulated example.
H-evidence plot.
Posterior intervals for $s(x_i)$ sorted by $\hat{s}(x_i)$.
The solid horizontal line is drawn at the estimate of $\sigma$ obtained
from fitting homoscedastic BART.
}\label{fig:oneDsim-check-heter}
\end{figure}

Figure \ref{fig:oneDsim-check-heter} 
displays the  inference for $s(x)$ in a way that will also
work for higher dimensional $x$.
We sort the observations according to the values of $\hat{s}(x)$.
We plot $\hat{s}(x)$ on the horizontal axis and posterior intervals for
$s(x)$ on the vertical axis.
The solid horizontal line is drawn at the estimate of $\sigma$ obtained
from the homoscedastic version of BART 
(see the top panel of Figure \ref{fig:assess-burn-in}).
We can clearly see that the posterior intervals are cleanly separated from the horizontal
line indicating that our discovery of heteroscedasticity is ``significant''.
In our real applications, where the units of $\sigma$ and $s(x)$ are the same as the units
of the response, we are able to assess the {\it practical significance} of the estimated
departure from constant variance.
We have found this plot useful in many applications.
We will use it in our other examples and call it the H-evidence plot.

\begin{figure}[ht!]
\begin{center}
\includegraphics[scale=.475]{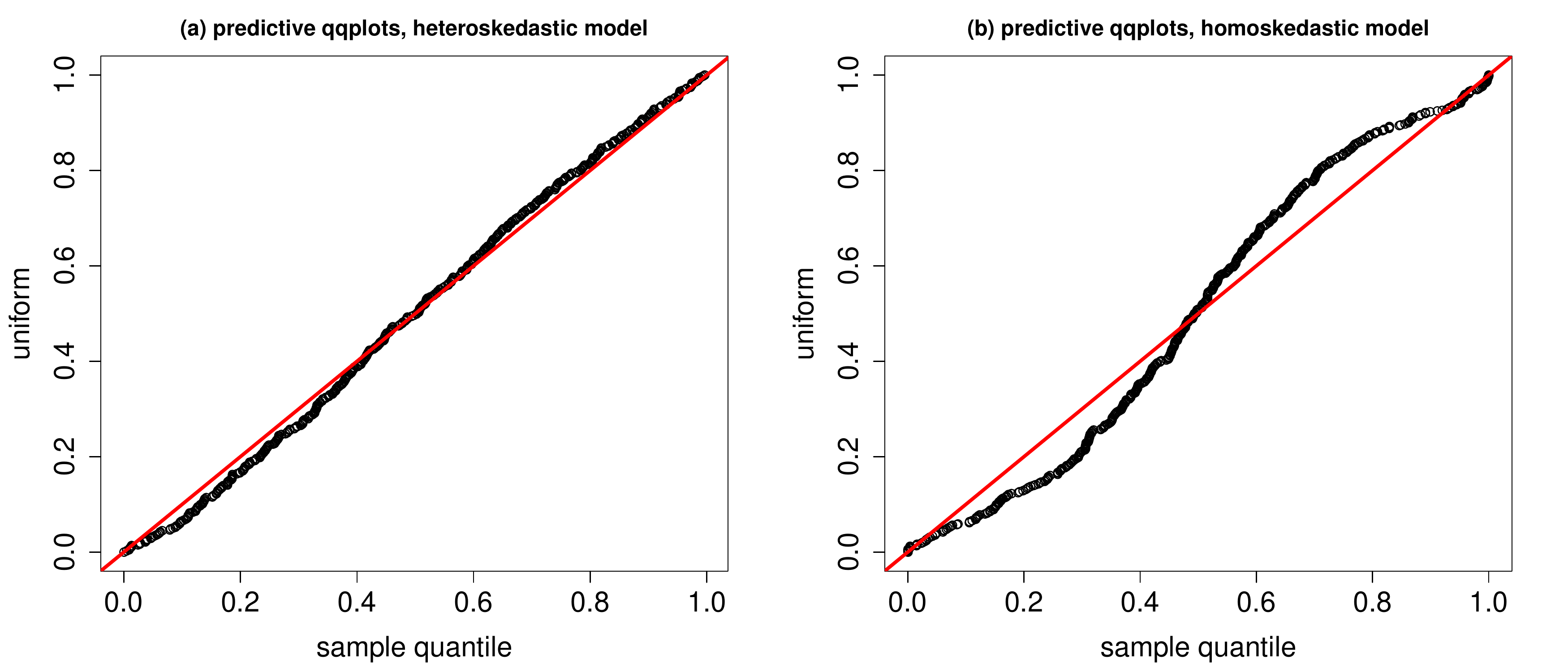}
\end{center}
\caption{
Simulated example.
Predictive qq-plots.
Left panel: HBART.
Right panel: BART.
}\label{fig:oneDsim-qq-plot}
\end{figure}

Figure \ref{fig:oneDsim-qq-plot} assesses the fit of our model
by looking at qq-plots (quantile-quantile plots) based on
the predictive distribution obtained from our model.
For each $i$ we obtain draws from $p(y \C x_i)$
and then compute the percentile of the observed $y_i$ in these draws.
If the model is correct, these percentiles should look like draws from
the uniform distribution on (0,1).
We use the qq-plot to compare these percentiles to draws from the uniform.  
Since the $(x_i,y_i)$ pairs
are from the test data so that we are evaluating the out-of-sample
predictive performance.

The left panel of
Figure \ref{fig:oneDsim-qq-plot}
shows the qq-plot obtained from our heteroscedastic model.
A ``45 degree'' line is drawn with intercept 0 and slope 1.
We see that our predictive percentiles match the uniform draws very well.
In the right panel, we do the same exercise, but this time our predictive 
draws are obtained from the homoscedastic model.
The failure of the homoscedastic BART model is striking.


Note that in our real applications we will be able to use the formats of 
Figures 
\ref{fig:oneDsim-check-heter}
and 
\ref{fig:oneDsim-qq-plot} 
to visualize the inference of HBART with high dimensional {\bf x}.

\subsection{Cars}\label{section:cars}

Perhaps one of the universally desired and undesired consumer expenditures in modern society is the 
purchase of a new or used car.  These large transactions are not only a challenge for consumers but also for the car sales industry itself.  Factors such as the commission-based nature of car sales, brand reliability, model reliability, warranty period, projected maintenance costs, and broader macro-economic conditions such as gasoline prices and job security all weigh on a consumers mind when purchasing a vehicle while trying to extract the most value for their dollar.  At the same time, these same variables weigh on car dealers as they try to extract the greatest profit from their inventory.  When important variables - such as gasoline prices - change, the effect can be profound.  For instance, if gasoline prices suddenly rise over a short time period, a consumer who recently purchased a large, expensive but fuel inefficient vehicle may find their economic assumptions change for the worse, while a dealership with a large inventory of such vehicles may suddenly be facing abnormally large losses rather than the normally small profits.

Studying pricing data for car sales is therefore a problem of great interest.  
However, besides understanding the mean behavior of price in response to changes in predictor variables, 
in such complex markets where profit margins are minimal, changes in the {\em variability} 
of prices could be equally important to understand for consumers and dealers alike.  
In this section, a sample of $n=1,000$ observations of used car sales data taken 
between 1994-2013 
is investigated.  
Notably, this dataset covers the 2007-2008 financial crisis and the subsequent recovery.  
The dataset consists of the response variable \texttt{price} (USD), 2 continuous predictor variables, 
\texttt{mileage} (miles) and \texttt{year}, and 4 categorical predictor variables, 
\texttt{trim}, \texttt{color}, \texttt{displacement} and \texttt{isOneOwner}.  
The categorical variables are summarized in Table \ref{tab:carscategorical}.

\begin{table}
\begin{tabular}{lll}
Variable & Levels & Description\\
\hline
\texttt{trim} & 430,500,550,other & Higher trim corresponds to higher-end vehicle.\\
\texttt{color} & black,silver,white,other & Color of vehicle.\\
\texttt{displacement} & 4.6, 5.5, other & Larger displacement corresponds to more powerful gas engine.\\
\texttt{isOneOwner} & true, false & Has vehicle had a single owner.\\
\hline
\end{tabular}
\caption{Summary of categorical predictor variables in the cars dataset.}
\label{tab:carscategorical}
\end{table}

\begin{figure}[ht!]
\begin{center}
\includegraphics[scale=.6]{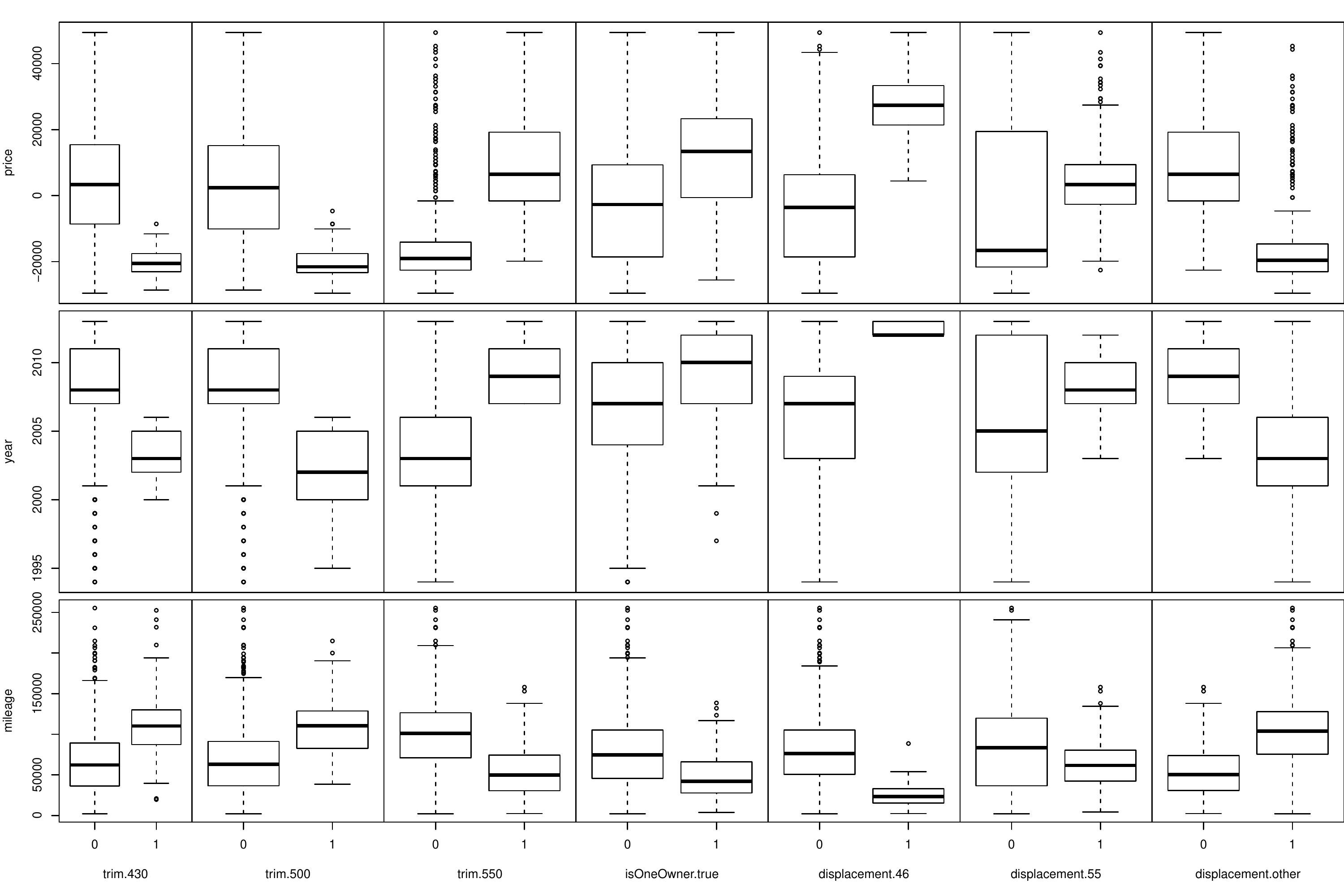}
\end{center}
\caption{
Used cars example.
Summary of response variable \texttt{price} and continuous predictors \texttt{mileage} and \texttt{year} by the levels of the important categorical predictors \texttt{trim}, \texttt{isOneOwner} and \texttt{displacement}.}
\label{fig:carsbycatvars}
\end{figure}

Expanding the categorical variables into binary dummy variables results in a total of 15 predictor variables.  
The relationship between active categorical predictors and the response variable 
\texttt{price} and continuous predictors \texttt{mileage} and \texttt{year} 
are summarized in Figure \ref{fig:carsbycatvars}.  
Note that the categorical predictor \texttt{color} does not appear in this 
figure as it has little marginal effect on the response.

This plot provides some notable summary information on how the categorical predictors marginally affect car price:
\begin{itemize}
\item \texttt{trim.430} and \texttt{trim.500} have lower median price, while \texttt{trim.550} has higher median price;
\item \texttt{displacement.46} and \texttt{displacement.55} has higher median price, while \texttt{displacement.other} has lower median price, but note that \texttt{displacement.55} is more or less located in the middle range of prices.
\end{itemize}

There is also evidence of collinearity between the categorical predictors and continuous predictor variables:
\begin{itemize}
\item  \texttt{trim.430} and \texttt{trim.500} have higher median mileage, while \texttt{trim.550} has lower median mileage;
\item \texttt{trim.430} and \texttt{trim.500} have lower median year (older cars), while \texttt{trim.550} has higher median year (younger cars);
\item \texttt{displacement.46} and \texttt{displacement.55} has lower median mileage and higher median year (younger cars) while \texttt{displacement.other} has higher median mileage/lower median year (older cars);
\item \texttt{isOneOwner.true} tends to correspond to younger cars with lower median mileage.
\end{itemize}

\begin{figure}[ht!]
\begin{center}
\includegraphics[scale=.6]{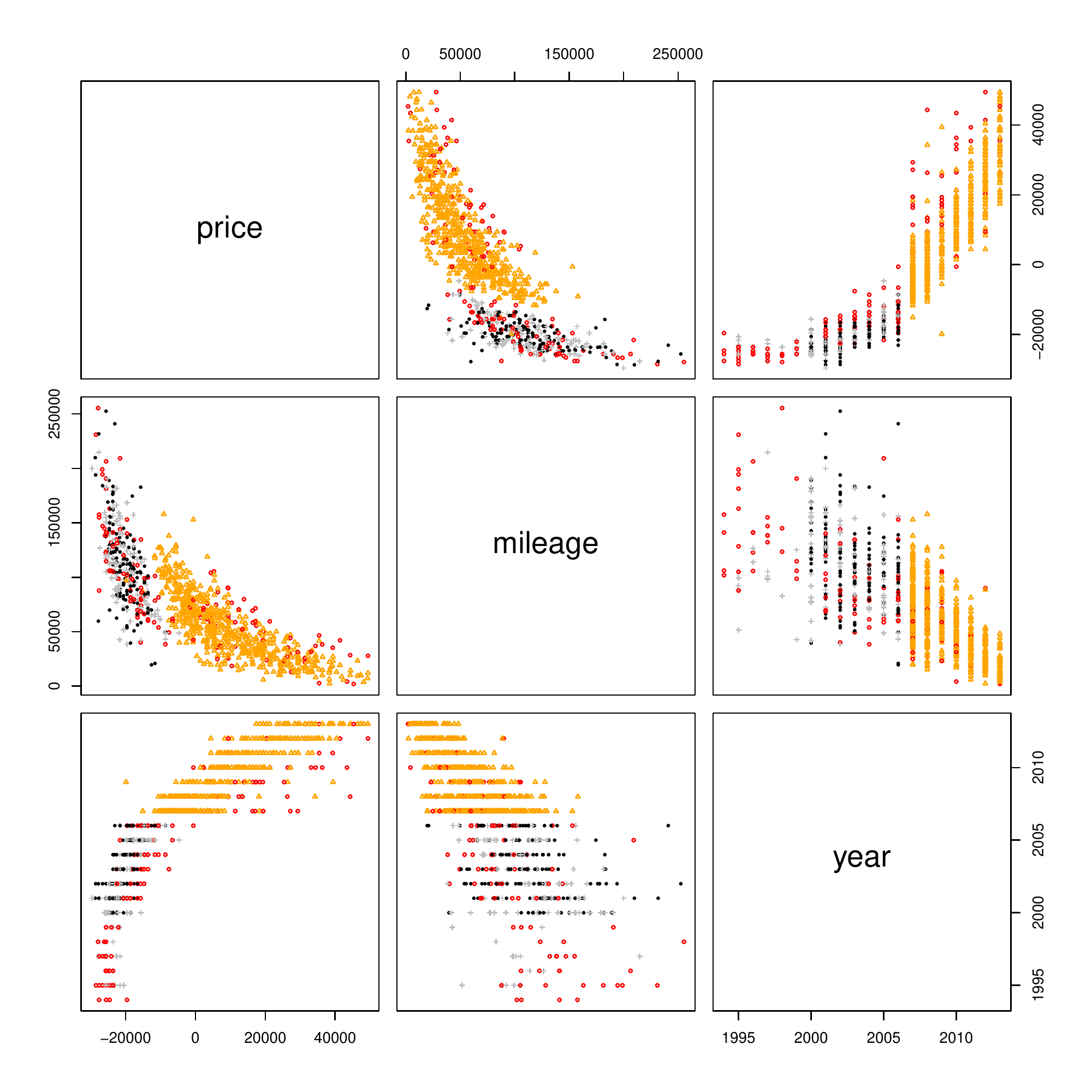}
\end{center}
\caption{
Used cars example.
Summary of continuous variables coded by level of \texttt{trim}.  \texttt{trim.430} shown by black solid dots, \texttt{trim.500} by grey `+', \texttt{trim.550} by orange triangles and \texttt{trim.other} by red `o'.}
\label{fig:carsbytrim}
\end{figure}

We can better understand these complex relationships by plotting the continuous variables color coded 
by each important categorical variable, such as shown in Figures \ref{fig:carsbytrim} to explore the effect of \texttt{trim}. 
This figures provide added insight.  
For instance, there is a clear curvilinear relationship between \texttt{mileage} 
and \texttt{price}, with higher \texttt{mileage} implying lower \texttt{price}.  
There is also a curvilinear relationship between \texttt{year} and \texttt{price} 
with higher \texttt{year} (younger car) implying higher \texttt{price}.  
However \texttt{mileage} and \texttt{year} are strongly negatively correlated 
(Pearson correlation of $-0.74$), so the amount of additional information for 
including one of these predictors after the other may be relatively small.

\begin{figure}[ht!]
\begin{center}
\includegraphics[scale=.6]{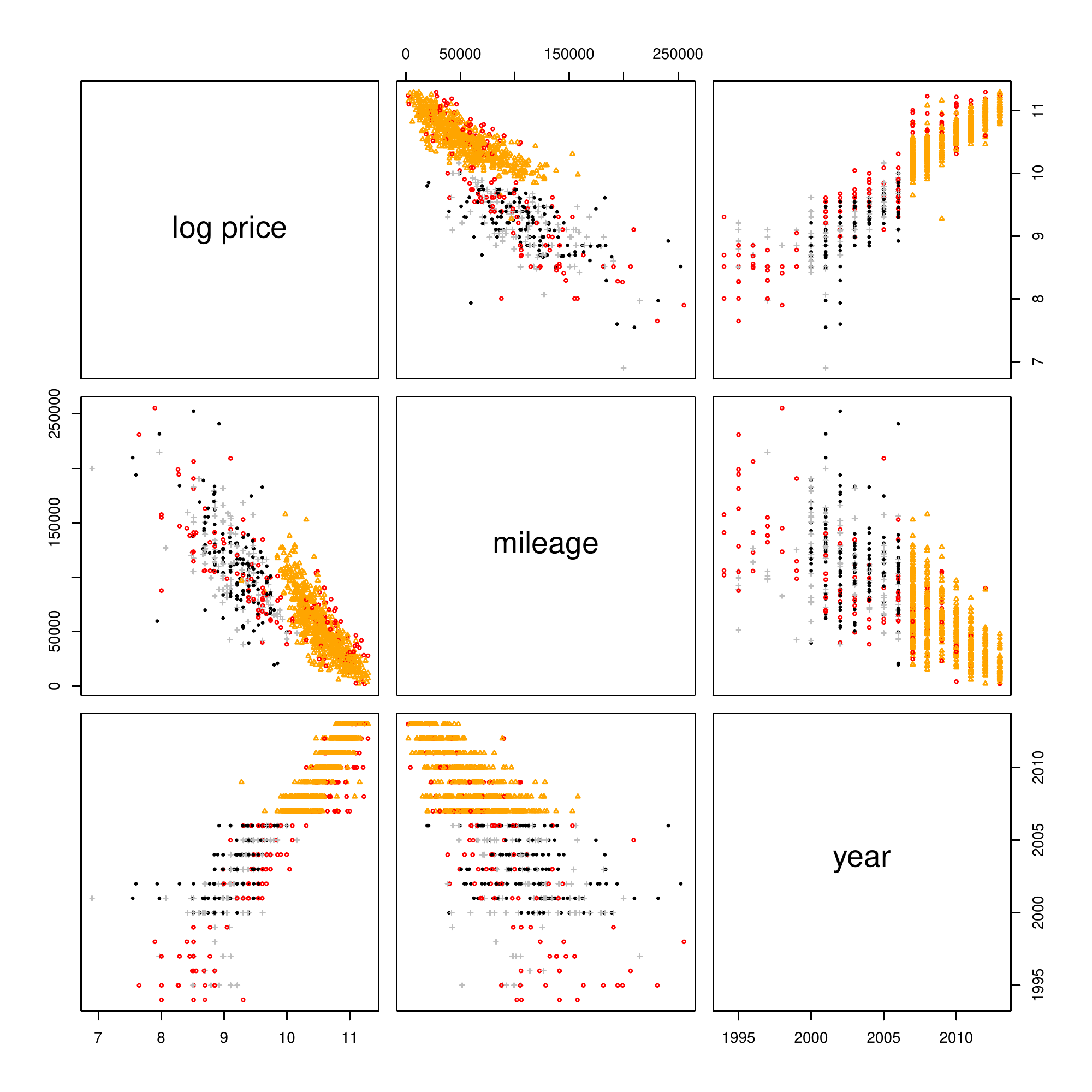}
\end{center}
\caption{
Used cars example.
Summary of log(\texttt{price}) and other continuous variables coded by level of \texttt{trim}.  \texttt{trim.430} shown by black solid dots, \texttt{trim.500} by grey `+', \texttt{trim.550} by orange triangles and \texttt{trim.other} by red `o'.}
\label{fig:carslogpricebytrim}
\end{figure}

Figure \ref{fig:carsbytrim} shows that  \texttt{trim.550} explains much of the ``jump'' 
seen in this figure for \texttt{year}$\geq 2007$, 
but not all as there are some \texttt{trim.other} cars that spread the entire range of years in the dataset.
There also appears to be a notable change in the spread of prices from 2007 onwards that 
does not seem likely to be explained by a mean function of the predictors.  
Rather, it suggests evidence for heteroscedasticity.  
It also appears that the spread changes in an abrupt manner around the year 
2007 across a wide range of price values, which suggests a simple approach such as 
log transforming the data to make the spread appear more constant might 
ignore potential insights that can be gained from analyzing this dataset.  
For instance, looking at the log transform of \texttt{price} coded by levels of \texttt{trim} 
shown in Figure \ref{fig:carslogpricebytrim}, 
a non-constant spread in \texttt{price} still appears evident across many years and many trim levels.  
Therefore, taking a simplistic transformation approach will not allow us to extract 
all the information available from the variables on their natural scale, and 
makes it more difficult to interpret the data.  
In addition, \cite{Box:Cox:1964} note that such power transformations 
alter both the variance and distribution of the data, 
making it difficult to separate second-moment corrections from higher-moment corrections, 
which aim to correct for normality rather than provide a concerted attempt at modeling and inferring heteroscedasticity.

Both variants of the BART model were applied to analyze this dataset.  When analyzing a real-world dataset, it is desirable to have a principled approach for tuning the settings of important hyperparameters.  In the case of the proposed HBART model, it is important to judiciously select the prior hyperparameter $\kappa$ in specifying the prior mean model. This parameter essentially controls whether BART will tend to fit smoother mean functions with a preference for greater heteroscedasticity (large $\kappa$) or fit more complex mean functions with a preference for homoscedasticity 
(small $\kappa$). However, selecting $\kappa$ using cross-validation based on MSE, for instance, is not adequate in our setting since we are interested in matching the first and second moments of our dataset rather than just the mean.   

Instead, we build on the idea of the qq-plots shown in section \ref{section:simulated} 
which did allow us to compare the fit of models from a {\em distributional} perspective. 
Rather than viewing a potentially large number of qq-plots, 
we use a measure of distance between distributions to compare percentiles calculated 
as in 
Figure \ref{fig:oneDsim-qq-plot}
to the uniform distribution using a 1-number summary.  
Our approach makes use of the so-called energy, or $e$-statistic proposed by \cite{Szekely:Rizzo:2004}, 
although many alternative metrics of distributional distance are available in the literature.

\begin{figure}[ht!]
\begin{center}
\includegraphics[scale=.8]{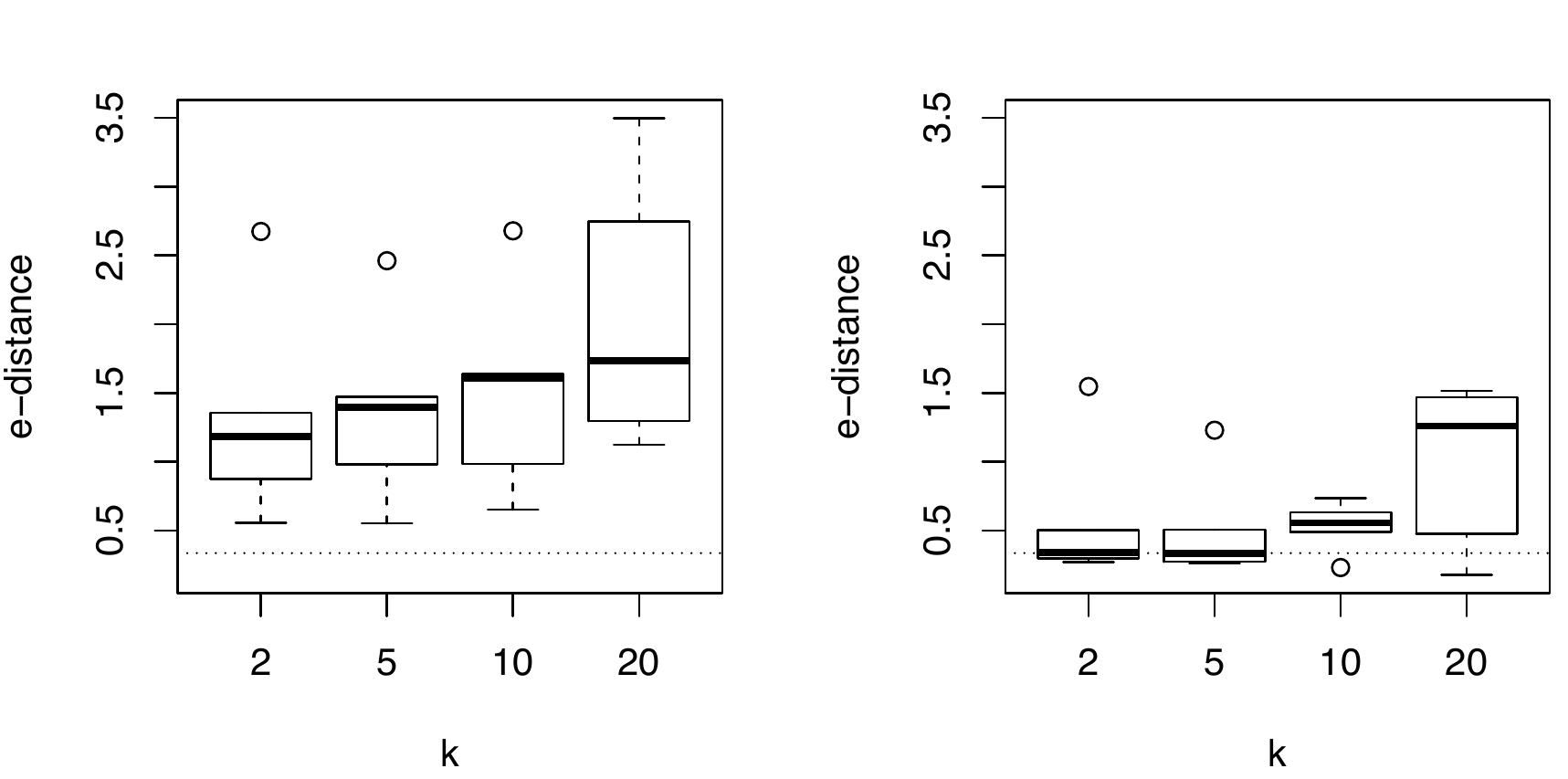}
\end{center}
\caption{
Used cars example.
Boxplots of $e$-distance from 5-fold cross-validation of BART (left pane) and HBART (right pane) in tuning the prior mean hyperparameter $\kappa$.  Horizontal dotted line corresponds to the median $e$-distance for $\kappa=5$ with the HBART model.}
\label{fig:carstunek}
\end{figure}

For both the usual homoscedastic BART model and the proposed heteroscedastic HBART model, 
5-fold cross-validation was performed to select the value of the hyperparameter 
$\kappa$ based on the $e$-statistic.  
That is, 
for each $(x,y)$ in the held-out fold,
we compute the percentile of $y$ in the predictive distribution of $Y \vert x$ computed from
the other folds.
We then use the $e$-statistic to compare these percentiles (one for each observation in the held-out fold)
to the uniform distribution.
This gives us one $e$-statistic comparison for each of the 5 folds.
This procedures is repeated for a small selection of plausible values of $\kappa.$
The result of this cross-validation is summarized in Figure \ref{fig:carstunek}.  
The cross-validation results suggest using 
a value of $\kappa=2$ (or smaller)
for the homoscedastic model while greater emphasis on smoothing the mean model 
is suggested with a cross-validated value of $\kappa=5$ for the heteroscedastic model.  
Besides the suggested settings for $\kappa$, note the overall much smaller values 
of $e$-distance for the heteroscedastic model, 
implying that this model is much better at capturing the {\em overall distributional pattern} 
of the observations rather than only the mean behavior.  
Generally, the $e$-statistic is much smaller for the heteroscedastic model.

\begin{figure}[ht!]
\begin{center}
\includegraphics[scale=1]{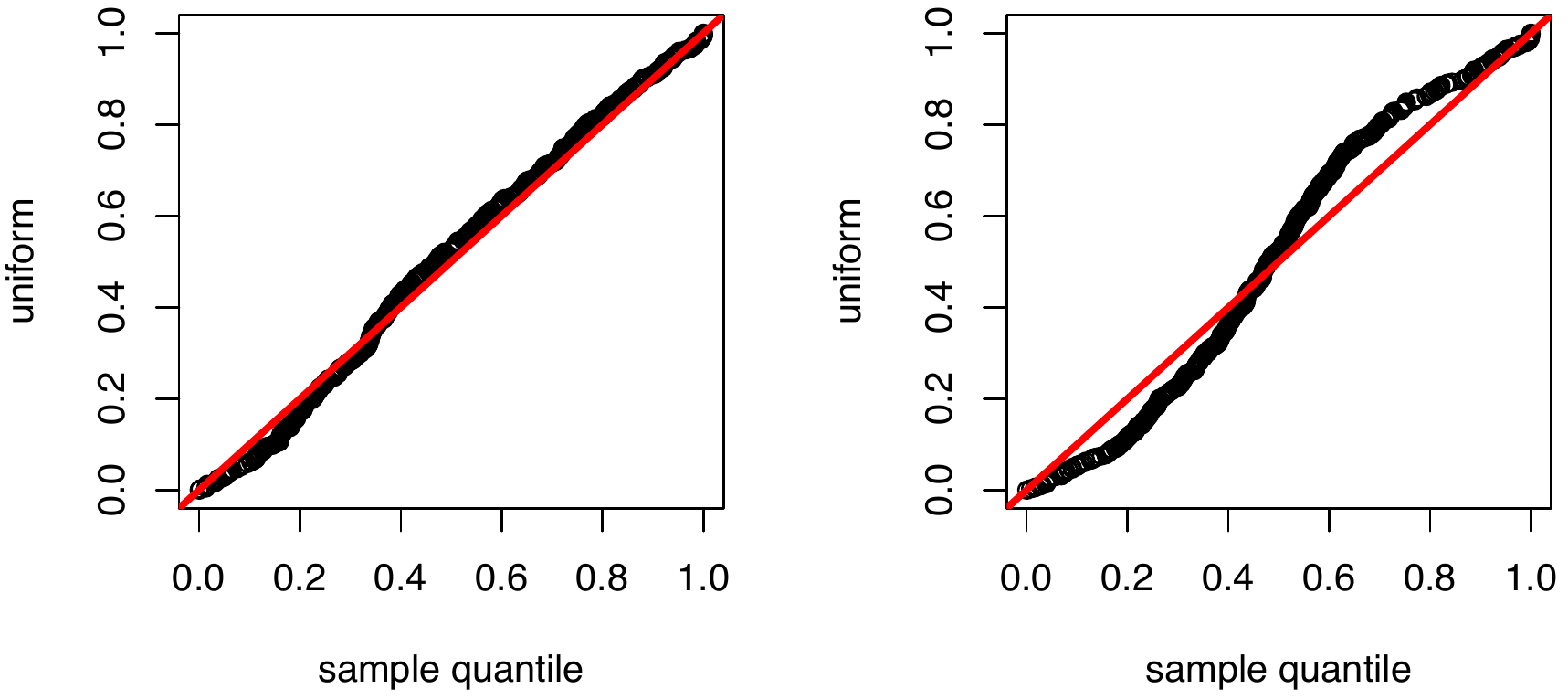}
\end{center}
\caption{
Used cars example.
Predictive qq-plot of posterior draws for \texttt{price} calibrated to the uniform distribution for the heteroscedastic model with $\kappa=5$ (left pane) and the homoscedastic model with $\kappa=2$ (right pane).}
\label{fig:carsqq}
\end{figure}

\begin{figure}[ht!]
\begin{center}
\includegraphics[scale=.6]{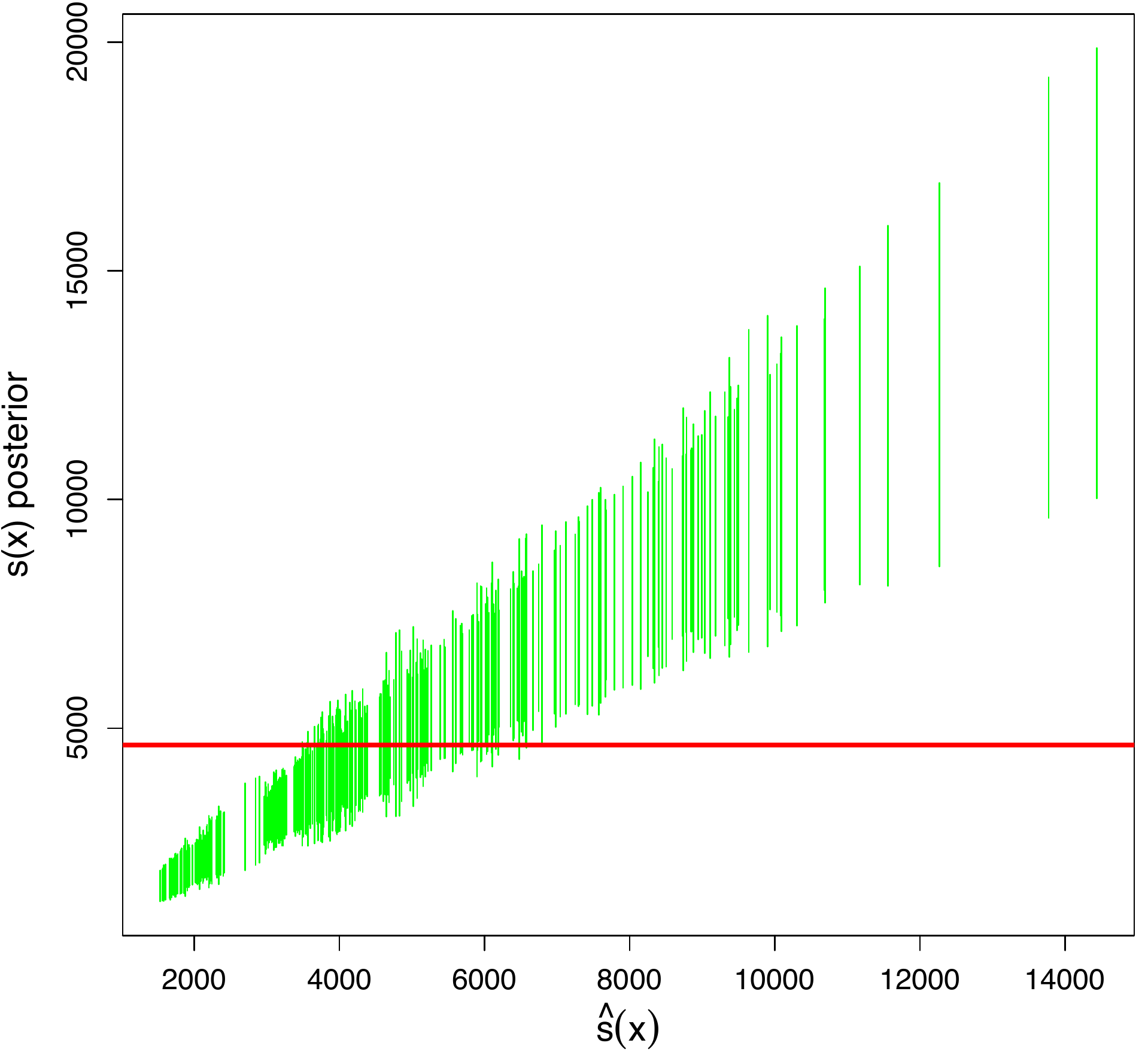}
\end{center}
\caption{
Used cars example.
H-evidence plot.
90\% posterior credible intervals for $s({\bf x})$ for the HBART model versus observation index sorted by level of the posterior mean standard deviation, $\hat{s}({\bf x})$.  The solid horizontal line shows the estimate of $\sigma$ from the  BART model for reference.}
\label{fig:carsheter}
\end{figure}


The corresponding predictive out-of-sample qq-plots for the homoscedastic model trained on 60\% of the full dataset with 
$\kappa=2$ and the heteroscedastic model with $\kappa=5$ are shown in Figure \ref{fig:carsqq}.  
This serves as an empirical graphical validation for the higher level of prior smoothing 
suggested by the cross-validated tuning of $\kappa$.  
The qq-plot for the heteroscedastic model is much closer to a straight line and dramatically better
than the qq-plot for the homoscedastic version.
The H-evidence plot in Figure \ref{fig:carsheter} serves as an additional check for evidence of heteroscedasticity: 
clearly, the posterior 90\% credible intervals do not cover the estimate of standard deviation 
from the homoscedastic model for a majority of the data.
Even  though there is considerable uncertainty about $s({\bf x})$ at the higher levels,
we have strong evidence that 
$s({\bf x}) > 10,000$
at the the higher levels and
$s({\bf x}) < 2,500$
at the lower levels.
These differences are  {\it practically} significant.

\begin{figure}[ht!]
\begin{center}
\includegraphics[scale=1]{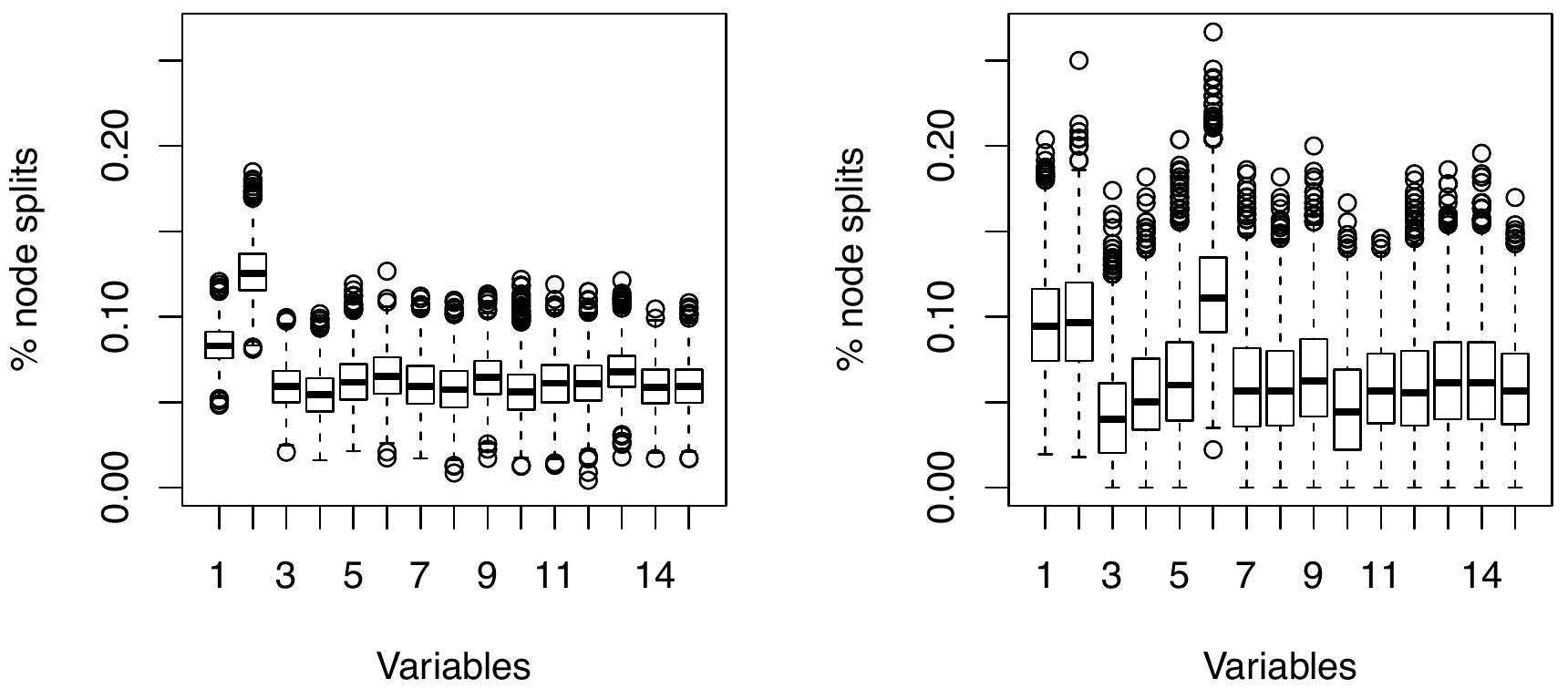}
\end{center}
\caption{
Used cars example.
Posterior variable activity from the HBART model for the mean (left pane) and $s({\bf x})$ (right pane).  The $x$-axis denotes the variable number while the $y$-axis is the posterior proportion of splits made on each variable.  Variables `1' and `2' correspond to \texttt{mileage} and \texttt{year} while variable `6' correponds to \texttt{trim.other}.}
\label{fig:carsvartivity}
\end{figure}

A large benefit of the proposed model is the ability to perform inference on both the mean {\em and} variance of the observed process.  By drawing from the posterior predictive of the $y$ process, one can learn what predictor variables affect changes in mean, variance or both.  For instance,
a standard measure of variable activity in BART is to calculate the percentage of internal tree node splits on each variable across all the trees.  With HBART, we can obtain this information for both the mean and standard deviation $s({\bf x})$ as shown in Figure \ref{fig:carsvartivity}.  These figures summarize some interesting findings.  The majority of tree splits for the mean model occur on the first two predictors in the dataset, \texttt{mileage} and \texttt{year}.  These two variables are also important in tree splits for the variance model as shown in the right pane.  As we suspected earlier, this includes the \texttt{year} variable which agrees with our exploratory plot in Figure \ref{fig:carslogpricebytrim} where a large jump in the spread of the response variable seems to occur from 2007 onwards.  \texttt{mileage} is also an important variable to model the spread of the data, which is not surprising since one would expect cars with similar characteristics with mileage near 0 may have well determined values while as the mileage increases the determination of value becomes more complex. Interestingly, a third variable seems strongly important for the variance trees while not particularly important for the mean trees: \texttt{trim.other}.  As shown in Figure \ref{fig:carslogpricebytrim}, cars with \texttt{trim.other} unusually span the entire range of years in the dataset, so it seems sensible that it may be an important variable in modeling the spread of the data.

\begin{figure}[ht!]
\begin{center}
\includegraphics[scale=1]{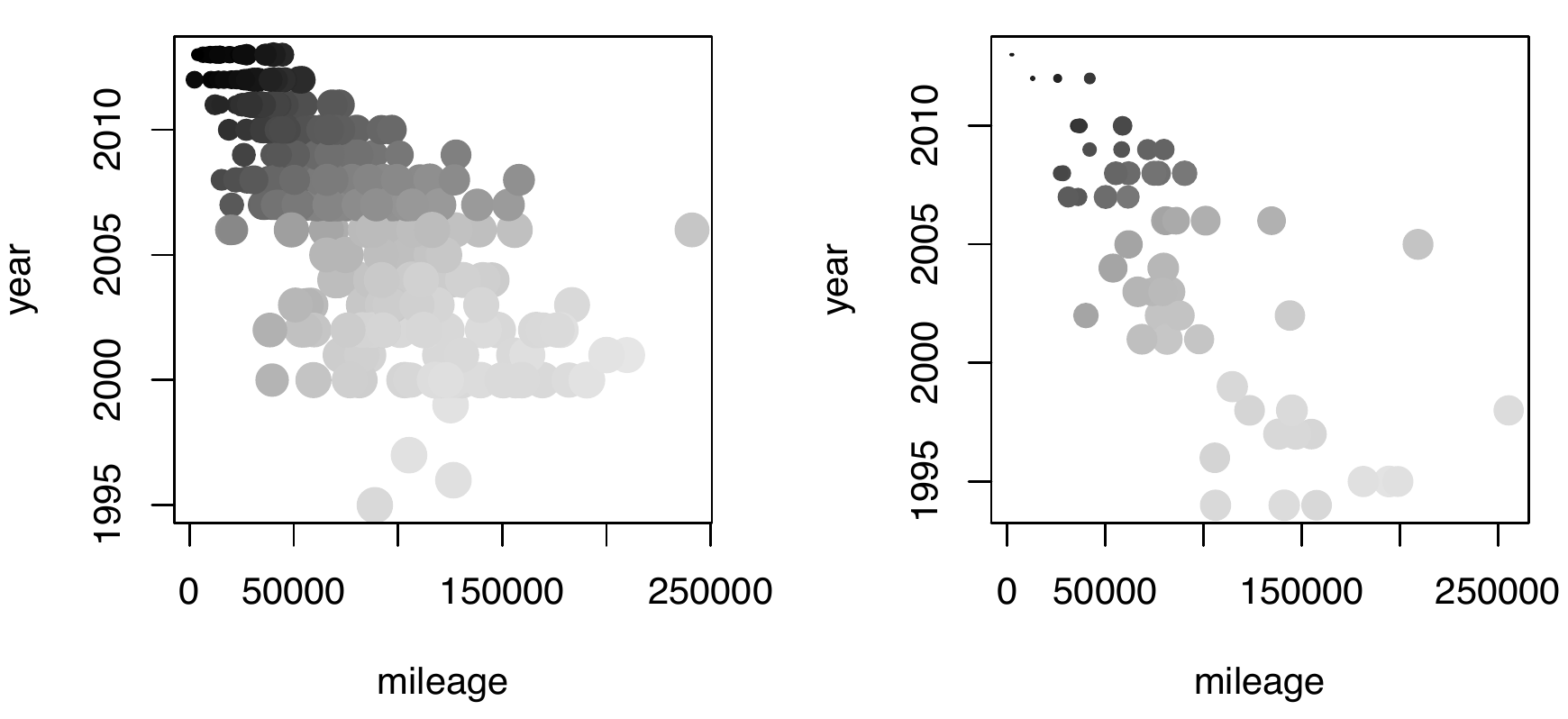}
\end{center}
\caption{
Used cars example.
Posterior median price for out-of-sample data plotted as a function of \texttt{year} and \texttt{mileage}.  The left pane corresponds to cars with \texttt{trim.other}=0 while the right pane corresponds to cars with \texttt{trim.other}=1.  The circles are shaded according to posterior predicted median \texttt{price} with lighter denoting lower \texttt{price} and darker denoting higher \texttt{price}.  Larger circles denote higher posterior predicted median standard deviation while smaller circles denote lower standard deviation. }
\label{fig:carsvarbytrim}
\end{figure}

Finally, Figure \ref{fig:carsvarbytrim} demonstrates how using the posterior samples from the fitted model can more clearly demonstrate the effect of \texttt{trim.other}.  This figure shows that as a function of \texttt{mileage} and \texttt{year}, the median predicted \texttt{price} displays the same general pattern for both levels of \texttt{trim.other}.  However, the variability becomes noticeably smaller for \texttt{trim.other}=1 at smaller \texttt{mileage} and more recent \texttt{year}.

\subsection{Fishery and Alcohol Examples}\label{section:Fish-Alc}
In this section we very briefly present results from two more examples.
In the first example the dependent variable $y$ is
the daily catch of fishing boats in the Grand Bank fishing grounds
\citep{fernandez:etal:2002}.
The explanatory ${\bf x}$ variables capture 
time, location, and characteristics of the boat.
After the creation of dummies for categorical variables, 
the dimension of ${\bf x}$ is 25.
In the second example, the dependent variable $y$ is
the number of alcoholic beverages consumed
in the last two weeks
\citep{kenkel:terza:2001}.
The explanatory ${\bf x}$ variables capture 
demographic and physical characteristics of the respondents as well as a 
key treatment variable indicating receipt of advice from a physician.
After the creation of dummies for categorical variables, 
the dimension of ${\bf x}$ is 35.

In both of the examples the response is constrained to be positive
and there is a set of observations with $y=0$
so that there is a clear sense in which our model
$Y = f({\bf x}) + s({\bf x}) Z$ does not account for these features of $Y$.
In both previous papers, careful modeling was done to capture the special nature
of the dependent variable.
Our interest here is to see how well our model can capture the data
given our flexible representations of $f$ and $s$ in the presence of 
a clear mispecification. 

\begin{figure}[ht!]
\begin{center}
\includegraphics[scale=.425]{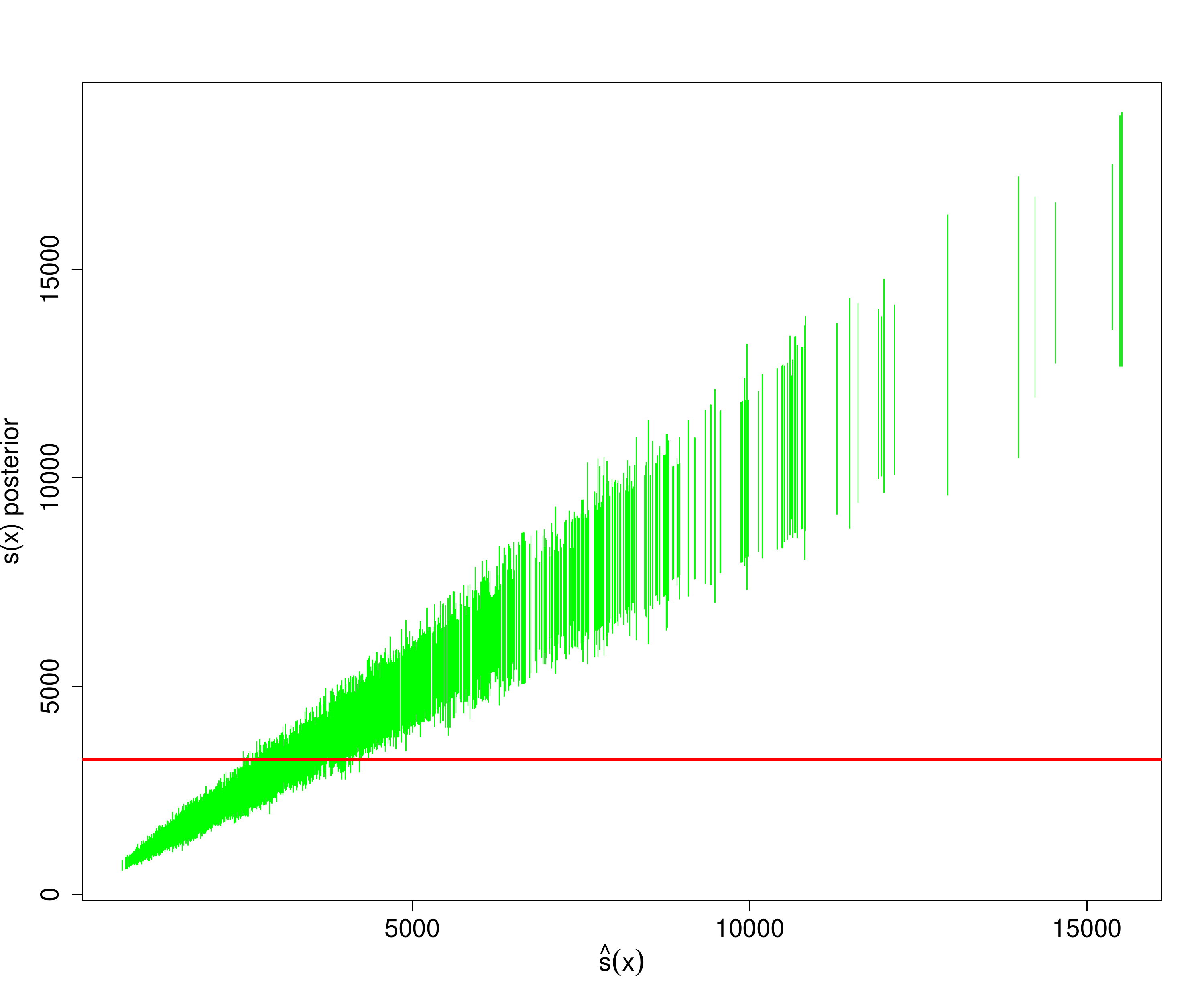}
\end{center}
\caption{
Fishery example.
H-evidence plot.
Posterior intervals for $s({\bf x}_i)$ sorted by $\hat{s}({\bf x}_i)$.
The solid horizontal line is draw at the estimate of $\sigma$ obtained
from fitting BART.
}\label{fig:Fishery_k2-check-heter}
\end{figure}

\begin{figure}[ht!]
\begin{center}
\includegraphics[scale=.475]{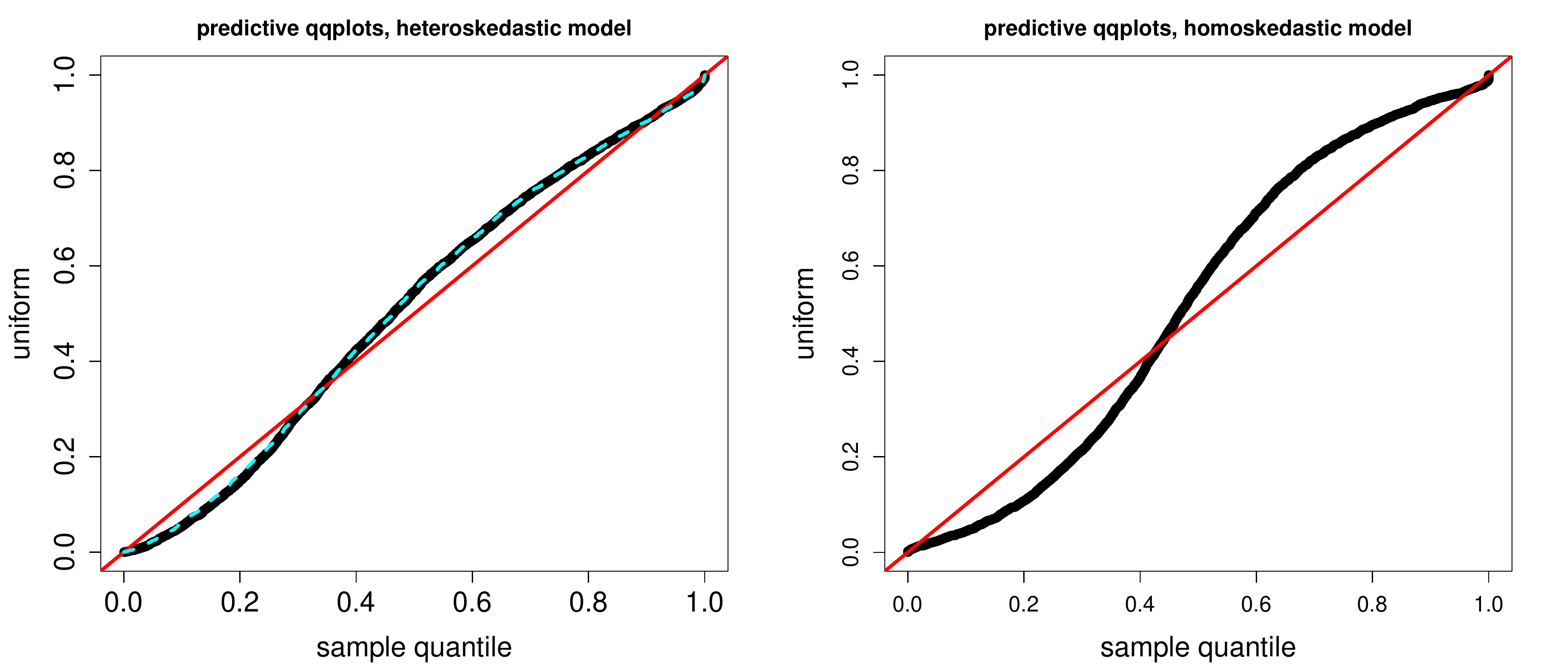}
\end{center}
\caption{
Fishery example.
Predictive qq-plots.
Left panel: HBART.
Right panel: BART.
}\label{fig:Fishery_k2-qq-plot}
\end{figure}

Figures
\ref{fig:Fishery_k2-check-heter} and \ref{fig:Fishery_k2-qq-plot}
present the results for the fish data using the same displays
we have employed in our previous examples.
In Figure \ref{fig:Fishery_k2-check-heter} we see very strong evidence of 
heteroscedasticity.  Our product of trees representation of $s$
enables the model to represent the data by being quite certain that
for some ${\bf x}$ the error standard deviation should be small.
Does this work?
In Figure
\ref{fig:Fishery_k2-qq-plot} we see the (in-sample) qqplots.
While the qqplot for the HBART model is not perfect,
it is a dramatic improvement over the homoscedastic fit and may be
sufficiently accurate for practical purposes.

In the left panel of Figure \ref{fig:Fishery_k2-qq-plot} we have also
plotted the qqplot obtained from the plug-in model
$Y \sim N(\hat{f}({\bf x}),\hat{s}({\bf x})^2)$.
This is represented by a dashed line.
It is difficult to see because it coincides almost exactly
with the qqplot plot obtained from the full predictive distribution.

Our feeling is that in many applications  
the representation
$Y \sim N(\hat{f}({\bf x}),\hat{s}({\bf x})^2)$
may be adequate and has an appealing simplicity.
Many users will be able to understand this output easily
without knowledge of the representations of $f$ and $s$.

\begin{figure}[ht!]
\begin{center}
\includegraphics[scale=.425]{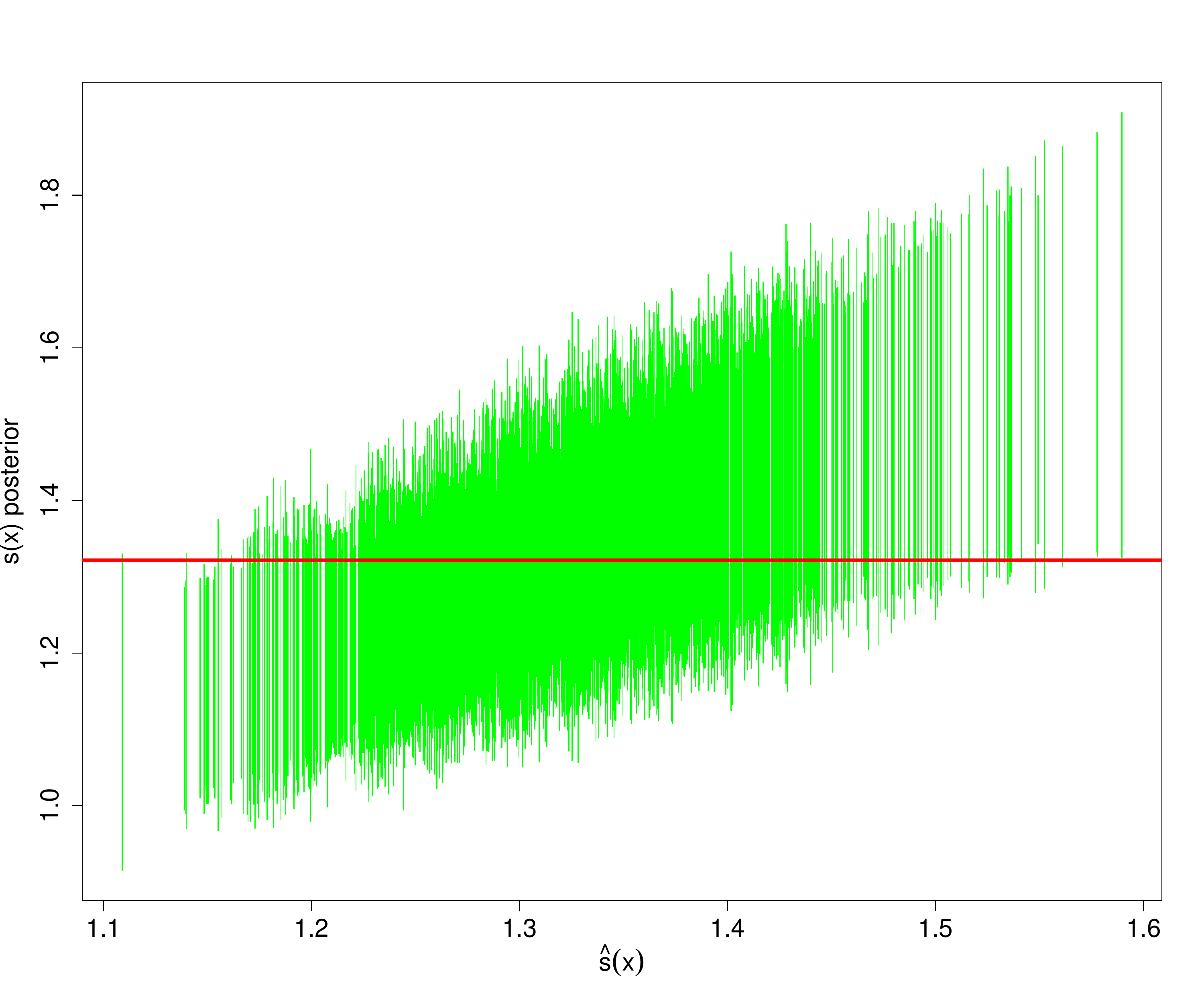}
\end{center}
\caption{
Alcohol example.
H-evidence plot.
Posterior intervals for $s({\bf x}_i)$ sorted by $\hat{s}({\bf x}_i)$.
The solid horizontal line is draw at the estimate of $\sigma$ obtained
from fitting BART.
}\label{fig:alcohol_k2-check-heter}
\end{figure}

\begin{figure}[ht!]
\begin{center}
\includegraphics[scale=.475]{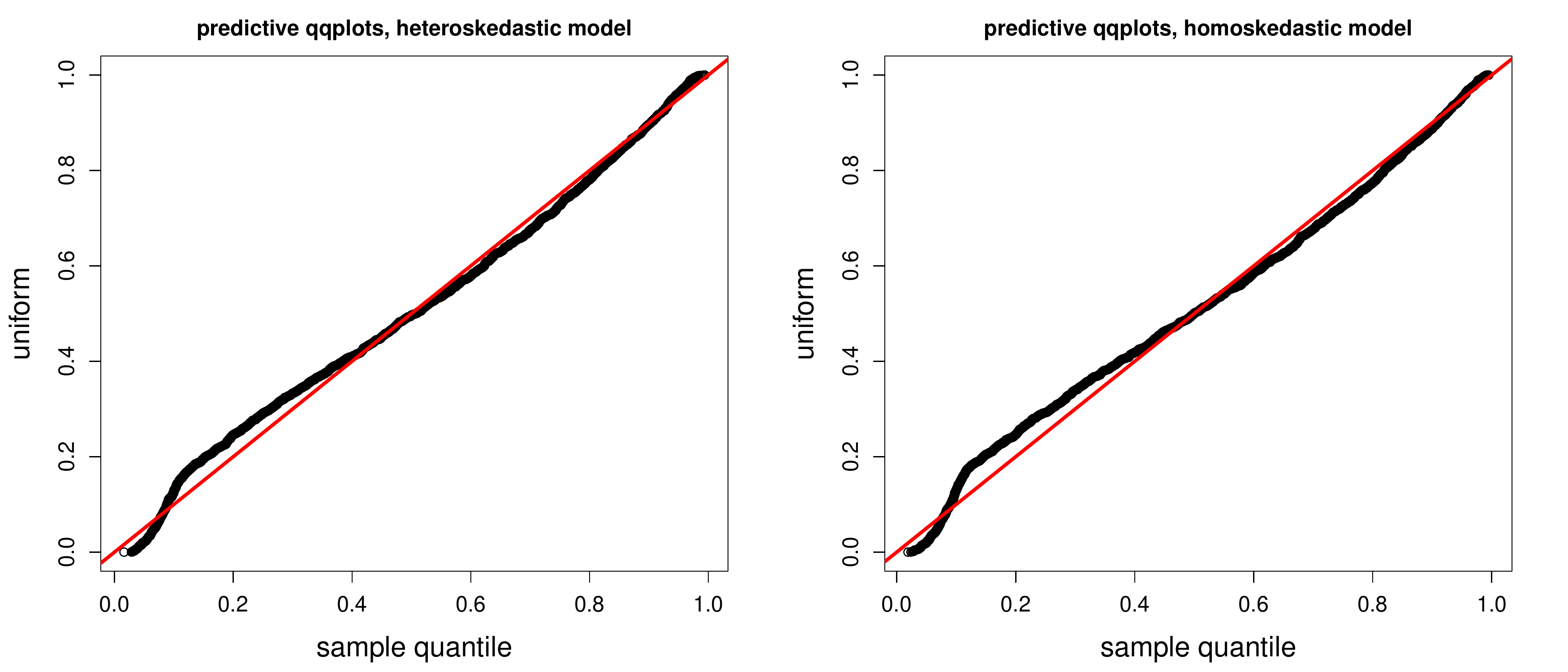}
\end{center}
\caption{
Alcohol example.
Predictive qq-plots.
Left panel: HBART.
Right panel: BART.
}\label{fig:alcohol_k2-qq-plot}
\end{figure}

Figures \ref{fig:alcohol_k2-check-heter} and \ref{fig:alcohol_k2-qq-plot}
give results for the Alcohol data again using the same format.
In this example the inference suggests that the homoscedastic version
is adequate and the (in-sample) qqplots are very similar.
In this case, even without the heteroscedastic model
the flexible $f$ captures the patterns reasonably well,
although the qqplots are not perfect.


\section{Conclusion}\label{section:conc}

In this paper we proposed a Bayesian non-parametric regression model that allows 
for flexible inference for both the conditional mean and variance using ensembles of trees. 
Modeling the mean takes the sum-of-trees approach made popular in the BART methodology while 
we introduced a product-of-trees ensemble for modeling the variance of the observed process.  
Our specification of the product-of-trees model gives us the conditional conjugacy needed for a
relatively simple MCMC algorithm. 
It also allows for an approach to prior specification which is 
no more complex to use than specifying priors in homoscedastic BART.

Our heteroscedastic model provides for a richer inference than standard BART.
In the context of our examples we developed tools for visualizing and tuning the inference.
The H-evidence plot allows us to quickly assess the evidence for predictor dependent variance.
The predictive qq-plot allows us to assess the full predictive distributional fit to the data.
While we often obtain good results from default prior specifications, it may be desirable
to tune the inference by using cross-validation for prior choice.
In our experience, calibrating the priors requires only cross-validating the hyperparameter `$\kappa$' 
in the prior mean model, thereby offering a fairly low level of complexity in terms of using the 
approach in real applied analyses.  
This is fairly surprising given the amount of flexibility that can be afforded by the model when the data demands it.
Rather than using mean square error to assess out-of-sample performance in cross-validation, 
we use the $e$-statistic to summarize the predictive qq-plot.

We demonstrated the proposed method on a simulated example, used cars sales data, a fisheries dataset and data on alcohol consumption.  The simulated example, where the true mean and variance functions were known, resulted in very accurate estimation while relying on nothing more than the default priors.  In the used cars data where a deeper analysis was performed, the model captures clear evidence of heteroscedasticity. 
Analysis of this data also revealed interesting differences in variable activity, where the trees for the mean had 2 highly active predictors while the trees for the variance had 3 highly active predictors.  The ability to extract such inferential information for both the mean and variance components of the data may have important practical consequences.

In the cars data, the qq-plot 
(Figure \ref{fig:carsqq})
indicates that our model has done a very effective job of capturing
the conditional distribution of sales price given the car characteristics.
In the Fishery example, the fit from the heteroscedastic model is
dramatically better than the homoscedastic fit
(Figure \ref{fig:Fishery_k2-qq-plot})
but not perfect.
In the alcohol data example
(Figures \ref{fig:alcohol_k2-qq-plot} and \ref{fig:alcohol_k2-check-heter})
we can easily infer that the heteroscedastic version does not improve
the fit. 

In the Fishery data example we also note
(Figure \ref{fig:Fishery_k2-qq-plot})
that the plug-in model
$Y \vert {\bf x} \sim N(\hat{f}({\bf x}),\hat{s}({\bf x})^2)$
does as well as the full predictive and may be an adequate approximation
to the conditional distribution.
In 
\cite{fernandez:etal:2002}
a particular feature of the data (a lump of responses at zero)
was noted and a model was developed to capture this feature.
Our model captures much of the pattern in the data 
and does so in a  way that is both automatic and simply interpretable.

The homoscedastic BART model 
$Y=f({\bf x}) + \sigma \, Z, \; Z \sim N(0,1)$ 
is clearly very restrictive
in modeling the error term.
Our heteroscedastic ensemble model
$Y=f({\bf x}) + s({\bf x}) \, Z, \; Z \sim N(0,1)$,
moves away from the BART model but still assumes an additive
error structure with conditionally Normal errors.
While we are currently working on ways to relax these assumptions,
there are  a great many ways in which the model may be elaborated.
Our feeling is that, in many applications, 
by focusing on the first and second moments,
the model presented in this paper
will capture much of the structure to be found in the data
in a relatively simple way.  Finally, we note that theoretical support for BART in the form of optimal posterior concentration rates has recently been established by \cite{rockova:pas:2017}, and anticipate that similar results can be obtained for HBART.  

%
%


\if0\blind
{
\section*{Acknowledgments}
This research was partially supported by the US National Science Foundation grants DMS-1106862, 1106974 and 1107046, the STATMOS research network on Statistical Methods in Oceanic and Atmospheric Sciences.  H.~A.~Chipman acknowledges support from the Natural Sciences and Engineering Research Council of Canada.  E.~I.~George acknowledges support from NSF grant DMS-1406563.  
}\fi

\clearpage



\bibliography{./heterobart.bib}
\bibliographystyle{./jasa}

\end{document}